# Fast switching dual Fabry-Perot-cavity-based optical refractometry for assessment of gas refractivity and density – estimates of its precision, accuracy, and temperature dependence

Martin Zelan,[1,a)] Isak Silander,[2] Thomas Hausmaninger,[2] and Ove Axner[2,a)]

[1]*Measurement Science and Technology, RISE Research Institutes of Sweden, SE-501 15 Borås, Sweden*
[2]*Department of Physics, Umeå University, SE-901 87 Umeå, Sweden*

Fabry-Perot-Cavity-based Optical Refractometry (FPC-OR) and Dual FPC-OR (DFCB-OR) have been shown to have excellent potential for characterization of gases, in particular their refractivity and density. However, their performance has in practice been found to be limited by drifts. To remedy this, drift-free DFPC-OR (DF-DFCB-OR) has recently been proposed. Accurate expressions for assessments of gas refractivity and density by DF-DFCB-OR as well as suggested methodologies for realization of a specific type of DF-DFCB-OR, termed Fast Switching DFCB-OR (FS-DFCB-OR), have been presented in two accompanying works. Based on these, this paper scrutinizes the performance and the limitations of both DF- and FS-DFCB-OR for assessments of refractivity and gas density, in particular their precision, accuracy, and temperature dependence. It is shown that both refractivity and gas density can be assessed by FS-DFCB-OR with a precision in the $10^{-9}$ range under STP conditions. It is demonstrated that the external (absolute) accuracy of FS-DFCB-OR is mainly limited by the accuracy by which the instantaneous deformation of the cavity or the higher order virial coefficients can be assessed. Since all major non-linear dependences of the DF-DFCB-OR signal have been identified, it is also shown that the internal accuracy, i.e. the accuracy by which the system can be characterized with respect to an internal standard, can be several orders of magnitude better than the absolute. Furthermore it is concluded that the temperature dependence of FS-DFCB-OR is exceptionally small, typically in the $10^{-8}$ to $10^{-7}$/°C range, and primarily caused by thermal expansion of the FPC-spacer material. Finally, this paper discusses means on how to design a FS-DFCB-or system for optimal performance and epitomizes the conclusions of this and our accompanying works regarding both DF- and FS-DFCB-OR in terms of performance and provides an outlook for both techniques. With this it is shown that our works can serve as a basis for future realizations of instrumentation for assessments of gas refractivity and density that can fully benefit from the extraordinary potential of FPC-OR. [Doc. ID XXXXX]

## I. INTRODUCTION

Optical Refractometry (OR) is a powerful tool for characterization of gases, with possible application ranging from pressure standards to detection of small changes in gas density.[1] An advantage of measuring density with OR is that the relation of the two, to first order, is independent of temperature.[2]

OR is normally performed by interferometry and the highest precision is achieved by the use of Fabry-Perot (FP) cavities where a measured change in frequency of a laser that is locked to a mode of the FP cavity can be associated with a change in refractivity.[3-13] FP-OR thus provides an important tool for accurate temperature independent characterization of gases under a wide variety of conditions, not least dynamic ones.

However, due to thermal expansion of various parts of the system, the technique is in reality not fully temperature independent, which, under certain conditions, can limit the performance of the system.[3, 7-10] A way to minimize the temperature influences is by the use of an additional reference cavity in such a way that the common mode temperature effects can be canceled or minimized. In this and two accompanying works[13, 14] this is referred to as Dual Fabry-Perot Cavity based Optical Refractometry (DFPC-OR).[3,7,9-12]

However despite the use of reference cavities, it has been shown by Egan et al, as well as by us,[7,9,12] that measurements are still affected by long term temperature drifts. Therefore, in order to utilize the full power of the technique, methodologies that allow for assessments of gas refractivity and density under drift-free conditions, referred to as Drift-Free (DF-DFPC-OR), are suggested and described in one of our accompanying works,[14] while a possible realization of this, termed Fast Switching DFCB-OR (FS-DFCB-OR) is described in Ref. [13].

In order to benefit the most from this new concept, we have in one of these works presented general and explicit expressions for the assessment of refractivity and gas density from measurements of the shift of the frequency of a laser field that is locked to a mode of a FP cavity (or alternatively, as the shift of the beat frequency measured from a mixing of two light fields on a common detector) as the cavity is evacuated.[14] The other one presents experimental results on which the DF-DFPC-OR concept are built and suggests some FS-DFPC-OR methodologies for how the technique best can be utilized for assessment of gas refractivity, density, and flows.[13]

In this paper we supplement the accompany works by scrutinizing the performance of the suggested DF-DFPC-OR technique and the FS-DFPC-OR methodologies in more detail. In particular, we investigate the precision, accuracy and residual temperature dependencies of the proposed FS-DFPC-OR methodology when gas refractivity and density are assessed.

Section 2 gives a short summary of the basis of the assessment of refractivity and gas density, based on Ref. [14]. Based on both an assessment of the experimental performance of a DFPC-OR instrumentation that is given in Ref. [13] and values from the literature, the sections 3 and 4 provide estimates of the precision and accuracy of assessment of the refractivity and gas density, respectively. Chapter 5 deals with the residual temperature dependence that show up in the assessment of refractivity and gas density due to second order effects, primarily thermal expansion of the cavity spacer material and molecular collision effects. Based on the conclusions from the sections 3 - 5, section 6 suggests means about how to characterize a FS-DFPC-OR system so that the accuracy of the technique can be significantly improved, by the use of external or alternately internal standards

---

a) Authors to whom correspondence should be addresses; electronic mail: martin.zelan@ri.se and ove.axner@umu.se

together with suitable characterization procedures. Finally, in the last section, we summarize and conclude shortly the results from analysis performed in this work together with our accompanying works.

## II. THE BASIS OF FAST SWITCHING FABRY-PEROT-BASED OPTICAL REFRACTOMETRY FOR ASSESSMENT OF REFRACTIVITY AND GAS DENSITY

### A. Assessment of refractivity

The basis of the DF-DFPC-OR technique and the principles of the FS-DFCB-OR methodologies are described in some detail in one of our accompanying papers.[14] It was shown that the refractivity of a gas, $n_i - 1$, can be assessed by FS-DFPC-OR, for the case with no relocking, by

$$n_i - 1 = \Omega(\varepsilon,\varsigma,\eta)\frac{\Delta \nu_l}{\nu_0}\left[1 + \Omega(\varepsilon,\varsigma,\eta)\frac{\Delta \nu_l}{\nu_0} + \left(\frac{\Delta \nu_l}{\nu_0}\right)^2\right]$$
$$+ (n_f - 1)\left[1 + 2\Omega(\varepsilon,\varsigma,\eta)\frac{\Delta \nu_l}{\nu_0} + 3\left(\frac{\Delta \nu_l}{\nu_0}\right)^2\right], \quad (1)$$

and, for the case with relocking, by

$$n_i - 1 = \frac{\Delta q}{q}\left[1 - \varepsilon\left(1 + \frac{\Delta q}{q}\right) + \varepsilon^2\right]$$
$$+ \frac{\Delta \nu_l}{\nu_0}\left(1 + \frac{\Delta q}{q}\right)\left[\Omega(\varepsilon,\varsigma,\eta) - 2\varepsilon(1-2\varepsilon)\frac{\Delta q}{q}\right]$$
$$+ \left(\frac{\Delta \nu_l}{\nu_0}\right)^2\left(1 + \frac{\Delta q}{q}\right)[2\Omega(\varepsilon,\varsigma,\eta) - 1] \quad (2)$$
$$+ (n_f - 1)\left(1 + \frac{\Delta q}{q}\right)\left[1 - 2\varepsilon\frac{\Delta q}{q} + 2\Omega(\varepsilon,\varsigma,\eta)\frac{\Delta \nu_l}{\nu_0}\right],$$

where $\Delta \nu_l$ is the change in frequency of laser light that is locked to a mode of the measurement cavity that takes place as the cavity is evacuated (which is assumed to be identical to the measured change in beat frequency), $\nu_0$ is the frequency of the cavity mode addressed in an empty cavity, and $\Omega(\varepsilon,\varsigma,\eta)$ represents $1 - \varepsilon + 2\varsigma + \eta + \varepsilon^2$, where $\varepsilon$ is a cavity deformation parameter, while $\varsigma$ and $\eta$ are parameters related to the dispersion of the mirrors and the gas, respectively, all properly defined in Ref. [14]. Moreover, $n_f - 1$ is the refractivity of the residual gas after the evacuation (to accommodate for the case that not all gas is removed from the cavity by the evacuation process, also referred to as the residual refractivity), $\Delta q$ represents the number of modes by which the laser field is shifted during the relocking process, while $q_0$ denotes the mode number of the cavity mode that the laser addresses when the cavity has been evacuated. The latter is, in turn, given by $\nu_0 / \nu_{FSR}^0$, where $\nu_{FSR}^0$ is the free spectral range of the empty cavity, given by $c / (2L_0)$, where $L_0$ is the length of the empty cavity. Everywhere, the subscripts $i$ and $f$ stand for initial and final conditions in the cell, corresponding to prior to and after the evacuation of the gas, respectively.

### B. Assessment of gas density

It was shown in the same paper[14] that the density of a gas in a cavity, prior to evacuation, $\rho_{n,i}$, can be assessed by the FS-DFPC-OR technique, for the case with no relocking, by

$$\rho_{n,i} = \rho_{n,f} + \frac{2}{3A_R}\tilde{\chi}_{\Delta \nu}\frac{\Delta \nu_l}{\nu_0}\left[1 + \tilde{B}_{\Delta \nu}\frac{\Delta \nu_l}{\nu_0} + \tilde{C}_{\Delta \nu}\left(\frac{\Delta \nu_l}{\nu_0}\right)^2 + ...\right], \quad (3)$$

and, for the case with relocking, by

$$\rho_{n,i} = \rho_{n,f} + \frac{2}{3A_R}\tilde{\chi}_{\Delta q}\left\{\frac{\Delta q}{q_0}\left[1 + \tilde{B}_{-}^{(\Delta q)^2}\frac{\Delta q}{q_0} + \tilde{C}_{-}^{(\Delta q)^3}\left(\frac{\Delta q}{q_0}\right)^2\right]\right.$$
$$+ \frac{\Delta \nu_l}{\nu_0}\left[\tilde{A}_{\Delta \nu}^{-} + \tilde{B}_{\Delta \nu}^{\Delta q}\frac{\Delta q}{q_0} + \tilde{C}_{\Delta \nu}^{(\Delta q)^2}\left(\frac{\Delta q}{q_0}\right)^2\right] \quad (4)$$
$$\left.+ \left(\frac{\Delta \nu_l}{\nu_0}\right)^2\left[\tilde{B}_{(\Delta \nu)^2}^{-} + \tilde{C}_{(\Delta \nu)^2}^{\Delta q}\frac{\Delta q}{q_0}\right]\right\},$$

where $\rho_{n,f}$ represents the residual density of the gas after the evacuation, henceforth referred to as the residual gas density, $A_R$ is the (number or molar) polarizability of the gas, $\tilde{\chi}_{\Delta \nu}$ is a dimensionless factor that is close to unity but includes minor contributions from $\varepsilon$, $\varsigma$, $\eta$, and $n_f - 1$, while $\tilde{B}_{\Delta \nu}$ and $\tilde{C}_{\Delta \nu}$ are normalized coefficients that, together with $\tilde{\chi}_{\Delta \nu}$, are given in Table 2 in Ref. [14]. For the case with relocking, $\tilde{\chi}_{\Delta q}$ is a dimensionless factor that is close to unity and solely includes contributions from $\varepsilon$ and $n_f - 1$, while $\tilde{B}_{-}^{(\Delta q)^2}$, $\tilde{C}_{-}^{(\Delta q)^3}$, $\tilde{A}_{\Delta \nu}^{-}$, $\tilde{B}_{\Delta \nu}^{\Delta q}$, $\tilde{C}_{\Delta \nu}^{(\Delta q)^2}$, $\tilde{B}_{(\Delta \nu)^2}^{-}$, and $\tilde{C}_{(\Delta \nu)^2}^{\Delta q}$ are normalized higher order coefficients that are given in Table 3 in the same reference, all expressed in terms of the $A_R$, $B_R$, and $C_R$ virial coefficients as well as the $\varepsilon$, $\varsigma$, and $\eta$ parameters. As above, all entities are properly defined in Ref. [14].

### C. Means of scrutinizing the FS-DFCB-OR signal

From the expressions given above, it is possible to estimate the precision and accuracy of the FS-DFPC-OR technique for assessment of both refractivity and gas density under a variety of conditions. It is also possible to assess the temperature dependence of the FS-DFCB-OR signals. However, although such estimates depend on how the experimental system is realized and which measurement conditions that are used, it is possible to draw some general conclusions about the performance of the technique. Such a scrutiny will be given here.

Moreover, although the evaluation is valid for DF-DFPC-OR in general, to couple to the methodologies presented in our accompanying paper,[13] we will in this work specifically consider the case with FS-DFPC-OR, and, in most cases, standard temperature and pressure condition (STP) conditions. Other situations can be handled is a similar manner.

## III. PRECISION AND ACCURACY OF FAST SWITCHING FABRY-PEROT-BASED OPTICAL REFRACTOMETRY FOR ASSESSMENT OF REFRACTIVITY

### A. Precision of the assessment of refractivity by FS-DFPC-OR

Precision is a measure of statistical variability, representing the degree to which repeated measurements under unchanged conditions show the same results, that can be seen as representing the resolution of an instrumentation. It can, for the types of systems considered there, be defined in two ways. It can either refer to

(i) the fluctuations of multiple assessments of the beat frequency prior to and after evacuation of a cavity exposed to a single gas evacuation, or

(ii) the fluctuations of a series of consecutive measurements of the shifts of the beat-note corresponding to a number of repeated gas evacuations.

Since the former refers to a single evacuation of the cavity, it will reflect the precision of the instrumentation excluding the influence of fluctuations of the gas evacuation. This will therefore here be referred to as the single gas evacuation (SGE) precision of the methodology. The latter definition refers to the precision obtained for a series of gas evacuations, thus including the gas evacuation process. This one will here be referred to as the multiple gas evacuation (MGE) precision.

Among the various entities on which the refractivity depends, as long as the instrumentation is run under drift-free conditions, the cavity deformation parameter, $\varepsilon$, as well as the dispersion parameters, $\varsigma$ and $\eta$, can be seen as being constant. This implies that also the $\Omega(\varepsilon,\varsigma,\eta)$ entity can be considered to be constant. Hence, none of these will contribute to the precision of the technique.

*1. Single gas evacuation precision*

This implies that the SGE precision of refractivity assessments, here denoted $\sigma_{n_i-1}^{SGE}$, and its relative counterpart, $\bar{\sigma}_{n_i-1}^{SGE}$, defined as $\sigma_{n_i-1}^{SGE}/(n_i-1)$, can be expressed as

$$\bar{\sigma}_{n_i-1}^{SGE} = \frac{\sigma_{n_i-1}^{SGE}}{n_i-1} = \sqrt{\bar{\sigma}_{\Delta\nu_l}^2 + \bar{\sigma}_{\nu_0}^2}, \quad (5)$$

where $\bar{\sigma}_{\Delta\nu_l}$ and $\bar{\sigma}_{\nu_0}$ are the individual contributions to the relative precision in the assessment of refractivity due to the relative fluctuations in $\Delta\nu_l$ and $\nu_0$, respectively, which, according to Eqs. (1) and (2), are given by

$$\bar{\sigma}_{\Delta\nu_l} = \frac{\Delta\nu_l/\nu_0}{n_i-1} \cdot \frac{\sigma_{\Delta\nu_l}}{\Delta\nu_l} = \frac{\sigma_{\Delta\nu_l}}{\Delta\nu_l} \quad (6)$$

and

$$\bar{\sigma}_{\nu_0} = \frac{\Delta\nu_l/\nu_0}{n_i-1} \cdot \frac{\sigma_{\nu_0}}{\nu_0} = \frac{\sigma_{\nu_0}}{\nu_0}, \quad (7)$$

where, in turn, $\sigma_{\Delta\nu_l}$ and $\sigma_{\nu_0}$ are the fluctuations in $\Delta\nu_l$ and $\nu_0$, and where the last step in the equations is valid solely for the case with no relocking. Note also that although the $(\Delta\nu_l/\nu_0)/(n_i-1)$ term is unity for the case with no relocking, it can be significantly smaller than this for the case with relocking.

Since all measurements of the beat frequency prior to the evacuation of the measurement cavity are made when the laser light is locked to a given cavity mode, $q_0$, and the same holds for the measurements performed after the evacuation (either $q_0$ or $q_0 + \Delta q$), neither $q_0$, nor $\Delta q$, will fluctuate during a single gas evacuation assessment. This explains why there are no contributions from any $\sigma_{\Delta q}$ and $\sigma_{q_0}$ term in Eq. (5).

Typical sizes of fluctuations affecting the precision of FS-DFCB-OR are compiled in Table 1, in which the data are taken from the FS-DFPC-OS system scrutinized in our accompanying paper,[13] for the case with no relocking and assuming a measurement time of 1 s. It can be concluded that the relative fluctuations in the laser frequency, $\sigma_{\nu_0}/\nu_0$, are significantly smaller than those in the shift of the frequency of the laser, $\sigma_{\Delta\nu_l}/\Delta\nu_l$. The $\bar{\sigma}_{\nu_0}$-term can therefore be neglected with respect to the $\bar{\sigma}_{\Delta\nu_l}$-term in Eq. (5). This implies that the relative fluctuations in the assessment of the refractivity by a single gas evacuation, $\bar{\sigma}_{n_i-1}^{SGE}$ is in practice given by $\bar{\sigma}_{\Delta\nu_l}$, which, for the system scrutinized in our accompanying paper,[13] thus is estimated to $2 \times 10^{-9}$. This implies that the SGE precision in the assessment of the refractivity, $\sigma_{n_i-1}^{SGE}$, under STP conditions, amounts to $5 \times 10^{-13}$.

Note that these numbers refer to the specific FS-DFPC-OS system evaluated in Ref. [13], and that this system has not been optimized for optimal performance for the measurement time considered here (1 s). Since it has been demonstrated that more stable systems can be realized, e.g. with a white-noise limited behavior to at least 100 s,[3] it is expected that some of the numbers given in Table 1 can be improved by construction of an instrumentation dedicated to FS-DFPC-OR (up to an order of magnitude).

Table 1. Fluctuations of various entities for the FS-DFCB-OR instrumentation scrutinized in Ref. [13], evaluated for a measurement time of 1 s.

| Entity | | Value |
|---|---|---|
| Fluctuations of the laser frequency | $\sigma_{\nu_0}$ | <100 Hz [a] |
| Fluctuations of the beat frequency | $\sigma_{\Delta\nu}$ | 100 Hz [a] |
| Fluctuations of the residual refractivity | $\sigma_{n_f-1}$ | $3 \times 10^{-12}$ [b] |
| Relative fluctuations of the laser frequency | $\dfrac{\sigma_{\nu_0}}{\nu_0}$ | $<5 \times 10^{-13}$ [c] |
| Relative fluctuations of the beat frequency | $\dfrac{\sigma_{\Delta\nu_l}}{\Delta\nu_l}$ | $2 \times 10^{-9}$ [d,e] |
| Relative fluctuations of the residual refractivity | $\dfrac{\sigma_{n_f-1}}{n_i-1}$ | $1 \times 10^{-8}$ [b,d] |

[a] Evaluated from Fig. 3 in Ref. [13].
[b] Estimated from the use of a pressure gauge with a precision of 1 mPa.
[c] At a wavelength of 1.5 μm.
[d] For nitrogen at STP conditions.
[e] Assuming a $\Delta\nu_l$ of 60 GHz, given by $\nu_0(n-1)$ under STP conditions.

*2. Multiple gas evacuation precision*

Multiple gas evacuation (MGE) measurements comprise a series of measurements for which the cell is being evacuated. The MGE precision, $\sigma_{n_i-1}^{MGE}$, can be seen as depending on partly on the same parameters as the SGE precision,[15] but also on the precision of the repeated gas evacuations. Hence, $\sigma_{n_i-1}^{MGE}$ [as well as its relative counterpart, $\bar{\sigma}_{n_i-1}^{ext}$, given by $\sigma_{n_i-1}^{ext}/(n_i-1)$] can be expressed as

$$\bar{\sigma}_{n_i-1}^{MGE} = \frac{\sigma_{n_i-1}^{MGE}}{n_i-1} = \sqrt{(\bar{\sigma}_{n_i-1}^{SGE})^2 + \bar{\sigma}_{n_f-1}^2}, \quad (8)$$

where $\bar{\sigma}_{n_f-1}$ is the contribution to the relative precision in the assessment of refractivity from the relative fluctuations in $n_f-1$ (i.e. the residual refractivity) from a series of consecutive evacuations of the cell, which, by the use of the Eqs (1) and (2), can be written as

$$\bar{\sigma}_{n_f-1} = \frac{\sigma_{n_f-1}}{n_i-1}, \quad (9)$$

where $\sigma_{n_f-1}$ represents the fluctuations of the residual refractivity from a series of gas evacuations, i.e. the fluctuations in $n_f-1$.

Since the latter is proportional to the fluctuations of the residual gas density in the cell, it is feasible to assume that, under fast switching conditions, unless no particular precautionary measures are taken, this will dominate over the SGE precision. In this case, the relative precision of the assessment of the MGE refractivity, $\bar{\sigma}_{n_i-1}^{MGE}$, will be limited by the relative fluctuations of the residual refractivity from a series of gas evacuations of the cell, i.e. $\bar{\sigma}_{n_f-1}$, whose value, in turn, depends on the gas system and the gas evacuation procedure.

*3. Multiple gas evacuation precision – With residual gas monitoring*

A means to alleviate these fluctuations is to assess the amount of gas in the cell after each evacuation by measuring the residual pressure, $p_f$. When the initial refractivity of the gas in the cell is to be assessed, the residual refractivity, $n_f-1$, can be estimated as

$$n_f - 1 = \frac{3A_R}{2}\rho_{n,f} = \frac{3A_R}{2}\frac{p_f}{RT}, \quad (10)$$

where $A_R$ and $R$ are the ordinary molar polarizability and the molar gas constant, respectively,[16] and fed back into the expressions used to assess the refractivity, i.e. either Eq. (1) or (2). This will reduce significantly the fluctuations of the assessment of the initial refractivity of the gas in the cell.

When the residual pressure is monitored, the relative fluctuations of the residual refractivity, can, by use of Eq. (6) in Ref. 14 and the ideal gas law, be estimated from

$$\bar{\sigma}_{n_f-1} = \frac{\sigma_{n_f-1}}{n_i-1} = \frac{n_f-1}{n_i-1} \frac{3A_R}{2(n_f-1)} \frac{\sigma_{p_f}}{RT} \approx \frac{p_f}{p_i} \frac{\sigma_{p_f}}{p_f}, \quad (11)$$

where $\sigma_{p_f}$ represents the fluctuations of the assessment of a small but fixed (residual) pressure, which, in practice, is given by the precision of the pressure gauge, while $p_i$ and $p_f$ are the pressure of the gas before and after the evacuations.

This shows that $\bar{\sigma}_{n_f-1}$ is not given by the resolution of the pressure gauge, which is defined as the ratio of its precision and the maximum pressure at which it has any response; it is, in general, significantly smaller (better) than this. It is given by the relative precision of the pressure gauge at the residual pressure, $\sigma_{p_f}/p_f$, reduced by a factor given by the degree of evacuation of the cell, given either by $(n_f-1)/(n_i-1)$ or $p_f/p_i$.[17] The relative fluctuations of the residual refractivity, $\bar{\sigma}_{n_f-1}$, can therefore in principle be several orders of magnitude below the resolution of the pressure gauge.

A comparison of the leading term of the relative SGE precision, i.e. $\bar{\sigma}_{\Delta v_l}$, and the relative precision of the residual refractivity, i.e. $\bar{\sigma}_{n_f-1}$, shows that the MGE precision of the assessment of the refractivity, i.e. $\sigma_{n_i-1}^{MGE}$, will be unaffected by the residual refractivity from a series of gas evacuations of the cell when the fluctuations of the pressure gauge, $\sigma_{p_f}$, is sufficiently small, i.e. whenever it fulfills

$$\sigma_{p_f} < \frac{2RT}{3A_R} \frac{\sigma_{\Delta v_l}}{v_0}. \quad (12)$$

For the case with $N_2$, for which $(2RT)/(3A_R)$ is $3.4 \times 10^8$ Pa, and for the system scrutinized in Ref. 13, for which, according to Table 1, $\sigma_{\Delta v_l}/v_0$ is $5 \times 10^{-13}$, the right hand side of this expression is 0.2 mPa. This implies that whenever $\sigma_{p_f}$ is below this, the relative fluctuations of the residual refractivity, $\bar{\sigma}_{n_f-1}$, will be smaller than those of the frequency shift of the laser, $\bar{\sigma}_{\Delta v_l}$, whereby they will not contribute noticeably to the MGE precision of the refractivity assessment, $\sigma_{n_i-1}^{MGE}$. When this is the case, the relative MGE precision of the assessment of refractivity by FS-DFPC-OR is equal to the relative SGE precision.

For the case when the residual pressure in the measurement cavity is monitored but when the condition in Eq. (12) is not fulfilled, the relative MGE precision of the assessment of refractivity by FS-DFPC-OR, $\bar{\sigma}_{n_i-1}^{MGE}$, will be given by $\bar{\sigma}_{n_f-1}$, which, under STP conditions, can be estimated to be $10^{-5} \sigma_{p_f}$.

## B. Accuracy of the assessment of refractivity by FS-DFPC-OR

### 1. General description

Accuracy, which often is interpreted as a measure of systematic errors or uncertainties, is frequently defined as the degree of closeness of a measurement (or a set of measurements) of a quantity to that quantity's true value, or the nearness of an assessment of an entity to its true value.[18]

The accuracy of FS-DFPC-OR for assessment of refractivity is here taken as the uncertainty of an assessment of the refractivity, here denoted $\delta_{n_i-1}$. As for precision, it is possible to conclude, from the Eqs. (1) and (2), that this entity, together with its relative counterpart, $\bar{\delta}_{n_i-1}$, given by $\delta_{n_i-1}/(n_i-1)$, can be expressed as

$$\bar{\delta}_{n_i-1} = \frac{\delta_{n_i-1}}{n_i-1} = \sqrt{\bar{\delta}_{\Delta v_l}^2 + \bar{\delta}_{v_0}^2 + \bar{\delta}_\varepsilon^2 + \bar{\delta}_\varsigma^2 + \bar{\delta}_\eta^2 + \bar{\delta}_{n_f-1}^2}, \quad (13)$$

where the $\bar{\delta}_{\Delta v}$, $\bar{\delta}_{v_0}$, $\bar{\delta}_\varepsilon$, $\bar{\delta}_\varsigma$, $\bar{\delta}_\eta$, and $\bar{\delta}_{n_f-1}$ are the individual contributions to the relative uncertainty in the assessment of refractivity due to uncertainties in the assessment of the shift of the laser frequency $\Delta v_l$, the laser frequency $v_0$, the cavity deformation parameter $\varepsilon$, the mirror dispersion parameter $\varsigma$, the gas dispersion parameter $\eta$, and the residual refractivity, $n_f-1$. As above, all entities are properly defined in Ref. 14. The various individual contributions to $\bar{\delta}_{n_i-1}$ are given by

$$\bar{\delta}_{\Delta v_l} = \frac{\Delta v_l / v_0}{n_i-1} \cdot \frac{\delta_{\Delta v_l}}{\Delta v_l} = \frac{\delta_{\Delta v_l}}{\Delta v_l}, \quad (14)$$

$$\bar{\delta}_{v_0} = \frac{\Delta v_l / v_0}{n_i-1} \cdot \frac{\delta_{v_0}}{v_0} = \frac{\delta_{v_0}}{v_0}, \quad (15)$$

$$\bar{\delta}_\varepsilon = \frac{\delta_\varepsilon}{1-\varepsilon} \approx \delta_\varepsilon, \quad (16)$$

$$\bar{\delta}_\varsigma = 2\frac{\Delta v_l / v_0}{n_i-1} \cdot \delta_\varsigma = 2\delta_\varsigma, \quad (17)$$

$$\bar{\delta}_\eta = \frac{\Delta v_l / v_0}{n_i-1} \cdot \delta_\eta = \delta_\eta, \quad (18)$$

and

$$\bar{\delta}_{n_f-1} = \frac{\delta_{n_f-1}}{n_i-1}, \quad (19)$$

respectively, where, as above, the last steps in the Eqs. (14), (15), (17), and (18) are valid solely for the case with no relocking. It is here assumed, as was discussed above, that neither $\Delta q$, nor $q_0$, will change between consecutive measurements, but also, as is discussed in some detail in Ref. 13, that they can be assessed without any uncertainty. This implies that none of them will contribute to the uncertainty in refractivity.

Estimated magnitudes of these uncertainties for a typical well-designed but uncharacterized FS-DFCB-OR system are compiled in Table 2. Data are taken from the literature as well as our accompanying paper[13] and are predominantly valid for the conditions considered in that work. The various uncertainties have been estimated as follows. The relative uncertainty of the beat frequency, $\delta_{\Delta v_l}/\Delta v_l$, has been taken as the performance of the particular frequency counter used in Ref. 13. The first value in the table ($10^{-6}$) corresponds to its unreferenced mode of operation, while the second one ($10^{-12}$) refers to the case when it is referenced to a GPS. The uncertainty of the laser frequency, $\delta_{v_0}$, (10 kHz) refers to the case when the laser is locked to a GPS referenced frequency comb, with an integration time of 1 s. The uncertainty of the cavity deformation parameter, $\delta_\varepsilon$, ($2 \times 10^{-5}/2 \times 10^{-4}$) is taken as 20% of the estimated value of $\varepsilon$ for both the closed type of cavity used in Ref. 14 (the first value) and an open type (the second value). The uncertainty of the cavity mirror dispersion parameter, $\delta_\varsigma$, ($2 \times 10^{-7}$) is likewise taken as 20% of the estimated value for $\varsigma$ for low dispersion mirrors and a 30 cm long cavity. The uncertainty of the gas dispersion parameter, $\delta_\eta$, ($1 \times 10^{-8}$) is solely taken as 1% of $\eta$ since gas dispersion assessments often can be done with an excellent level of uncertainty. Finally, the uncertainty of the residual refractivity, $\delta_{n_f-1}$, ($6 \times 10^{-11}$) is estimated from assessments of the pressure using nitrogen gas and a pressure gauge with an accuracy of 20 mPa.

### 2. Accuracy of an uncharacterized system

For illustrational purposes, these individual contributions to the uncertainty of the residual refractivity are schematically displayed as a function of gas density (and pressure) in Fig. 1a and under STP conditions in Fig. 1b.

Table 2. Uncertainties in various entities for a typical uncharacterized FS-DFCB-OR instrumentation for assessment of refractivity. Data taken from the literature. Parameters and entities are defined in Ref. [14].

| Entity | | Value |
|---|---|---|
| Relative uncertainty of the beat frequency | $\dfrac{\delta_{\Delta\nu_l}}{\Delta\nu_l}$ | $10^{-6}/10^{-12\,a}$ |
| Uncertainty of the laser frequency | $\delta_{\nu_0}$ | 10 kHz [b] |
| Relative uncertainty of the laser frequency | $\dfrac{\delta_{\nu_0}}{\nu_0}$ | $5 \times 10^{-11\,c}$ |
| Uncertainty of the cavity deformation parameter | $\delta_\varepsilon$ | $2 \times 10^{-5}/2 \times 10^{-4\,d}$ |
| Uncertainty of the cavity mirror dispersion parameter | $\delta_\varsigma$ | $2 \times 10^{-7\,c,d,e,f}$ |
| Uncertainty of the gas dispersion parameter | $\delta_\eta$ | $1 \times 10^{-8\,d,g,h}$ |
| Uncertainty of the residual refractivity | $\delta_{n_f-1}$ | $6 \times 10^{-11\,h,i}$ |
| Relative uncertainty of the residual refractivity | $\dfrac{\delta_{n_f-1}}{n_i-1}$ | $2 \times 10^{-7\,h,i,j}$ |

[a] Agilent 53230A, time base 1-year, First value: unreferenced mode of operation, second value: referenced to GPS.[19]
[b] Relative to a GPS referenced frequency comb, measured over 1 s.[8]
[c] At a wavelength of 1.5 μm.
[d] Taken as 20% of the cavity deformation parameter, $\varepsilon$. First value: for a well-constructed closed cavity, second value: for an open cavity.[14]
[e] Taken as 20% of the mirror dispersion parameter, $\varsigma$, for low dispersion mirrors discussed in Ref. [14].
[f] For a cavity length of 30 cm.
[g] Taken as 1 % of the gas dispersion parameter, $\eta$, as discussed in Ref. [14].
[h] For nitrogen.
[i] Using a well-calibrated pressure gauged, with an accuracy of 20 mPa [e.g. Ceravac, CTR 100, 1 Torr, Oerlicon].
[j] Under STP conditions.

Figure 1a shows that the contribution to the relative uncertainty of the refractivity from the relative uncertainty of the residual refractivity decreases with the gas density to be assessed, $\rho_{n,i}$, (lower x-axis) and with the corresponding pressure, $p_i$, (upper x-axis). The contributions from the relative uncertainties of the other entities are independent of this density (pressure). Of the others, the contribution from the relative uncertainty in the instantaneous change in length of the cavity (due to the presence of the gas in the cavity), $\bar{\delta}_\varepsilon$, dominates over the others. This implies that, for the conditions considered, for high densities (pressures), above 0.04 mol/m³ (~ 0.1 kPa) for open cavities and above 0.4 mol/m³ (~ 1 kPa) for closed ones, the major contribution to the relative uncertainty in refractivity is that from the instantaneous change in length of the cavity. For densities (pressures) below these, it is mainly dominated by the uncertainty in the residual refractivity.

Figure 1(b), which shows the successive order of importance for all entities considered above under STP conditions, displays that the uncertainty of the assessment of refractivity by FS-DFPC-OR is assumed to be primarily given by the uncertainty of the cavity deformation, $\bar{\delta}_\varepsilon$, (in agreement with the findings in Refs. [7, 9]) while those of the other entities plays a smaller role. For conditions that differ from those considered, e.g. for the case with relocking, the relative order of the contributions of the various entities can differ.

This shows that an uncharacterized FS-DFCB-OR system cannot be expected to provide a relative accuracy for assessment of refractivity that is better than in the $10^{-5}$ - $10^{-4}$ range. However, as is discussed in more detail below, if the system could be characterized by one type of gas whose refractivity is well-known (i.e. known with a low uncertainty) under at least one set of conditions, the system would demonstrate a significantly better accuracy.

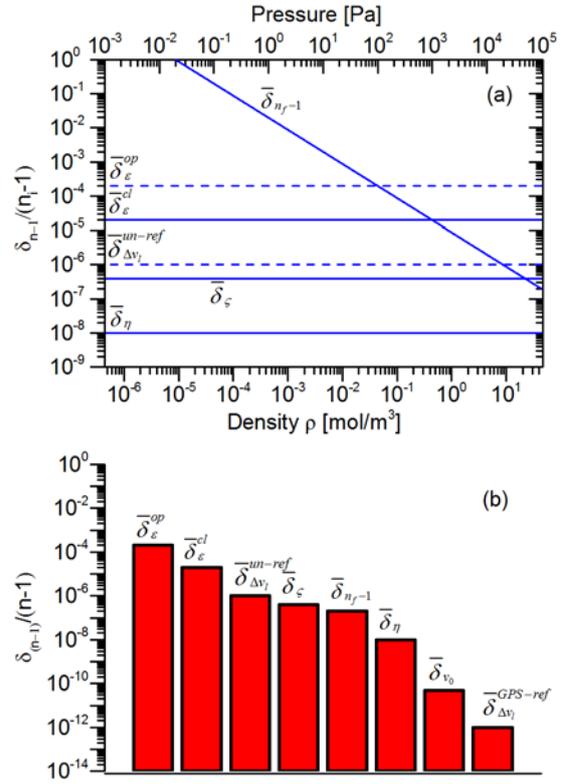

FIG 1. Panel (a) Schematic illustration of the relative contributions to the uncertainty of the refractivity, $\bar{\delta}_{n_i-1}$, expressed as $\delta_{n_i-1}/(n_i-1)$, from various physical entities in terms of gas density to be assessed, $\rho_{n,i}$, (lower x-axis). For clarity, the corresponding pressure, $p_i$, is given by the upper x-axis. The horizontal curves represent, from top to bottom, the relative contributions from the residual refractivity ($\bar{\delta}_\varepsilon^{op}$ and $\bar{\delta}_\varepsilon^{cl}$, open and closed cavity, respectively), the shift of the laser frequency for an unreferenced mode of operation of the frequency counter, $\bar{\delta}_{\Delta\nu_l}^{un-ref}$, mirror dispersion, $\bar{\delta}_\varsigma$, and gas dispersion, $\bar{\delta}_\eta$, (where the latter two represent the case with no relocking) respectively. The slanted solid curve corresponds to the contribution from the residual refractivity, $\bar{\delta}_{n_f-1}$. Panel (b) Schematic illustration of the corresponding situation under STP conditions, including also the contributions from the laser frequency, $\bar{\delta}_{\nu_0}$, and the shift of the laser frequency for the case with a frequency counter that is referenced to a GPS, $\bar{\delta}_{\Delta\nu_l}^{GPS-ref}$. In both panels, $\bar{\delta}_{\Delta\nu_l}^{GPS-ref}$ and $\bar{\delta}_{\Delta\nu_l}^{un-ref}$ refer to the cases when the frequency counter is unreferenced or referenced with respect to a GPS, respectively, while $\bar{\delta}_\varepsilon^{cl}$ and $\bar{\delta}_\varepsilon^{op}$ correspond to the cases when a closed and an open type of cavity is used, respectively.

## IV. PRECISION AND ACCURACY OF FAST SWITCHING FABRY-PEROT-BASED OPTICAL REFRACTOMETRY FOR ASSESSMENT OF GAS DENSITY

### A. Precision of the assessment of gas density by FS-DFPC-OR

*1. Single gas evacuation precision*

Since none of the virial coefficients, and thereby none of the higher order coefficients in the expression for the gas number density in Eq. (3), contribute to the precision of the assessment, the SGE precision of the assessment of the gas density by the FS-DFCB-OR technique, here denoted $\sigma_{\rho_{n,i}}^{SGE}$, and its relative counterpart, $\bar{\sigma}_{\rho_{n,i}}^{SGE}$, defined as $\sigma_{\rho_{n,i}}^{SGE}/\rho_{n,i}$, can be expressed as

$$\bar{\sigma}_{\rho_{n,i}}^{SGE} = \frac{\sigma_{\rho_{n,i}}^{SGE}}{\rho_{n,i}} = \sqrt{\bar{\sigma}_{\Delta\nu_l}^2 + \bar{\sigma}_{\nu_0}^2} \;, \qquad (20)$$

where $\bar{\sigma}_{\Delta\nu_l}$ and $\bar{\sigma}_{\nu_0}$ now are the individual contributions to the relative precision (i.e. the relative fluctuations) in the assessment of gas density due to the relative fluctuations in $\Delta\nu_l$ and $\nu_0$, respectively, which, according to Eqs. (3) and (4), are given by

$$\bar{\sigma}_{\Delta v_l} = \frac{\frac{2}{3A_R}\tilde{\chi}_{\Delta q}}{\rho_{n,i}} \cdot \frac{\sigma_{\Delta v_l}}{v_0} = \frac{\Delta v_l / v_0}{n_i - 1} \cdot \frac{\sigma_{\Delta v_l}}{\Delta v_l} = \frac{\sigma_{\Delta v_l}}{\Delta v_l} \qquad (21)$$

and

$$\bar{\sigma}_{v_0} = \frac{\frac{2}{3A_R}\tilde{\chi}_{\Delta q}\frac{\Delta v_l}{v_0}}{\rho_{n,i}} \cdot \frac{\sigma_{v_0}}{v_0} = \frac{\Delta v_l / v_0}{n_i - 1} \cdot \frac{\sigma_{v_0}}{v_0} = \frac{\sigma_{v_0}}{v_0}, \qquad (22)$$

respectively, where again the last steps are valid solely for the case with no relocking. This shows that the relative SGE fluctuations of an assessment of gas density by FPC-OR, $\bar{\sigma}_{\rho_{n,i}}^{SGE}$, are given by the same relative entities as when refractivity is assessed. Hence, since it was argued above that $\bar{\sigma}_{v_0}$ is significantly smaller than $\bar{\sigma}_{\Delta v_l}$, it can be concluded that also $\bar{\sigma}_{\rho_{n,i}}^{SGE}$ is primarily given by $\bar{\sigma}_{\Delta v_l}$. Moreover, since it was also found that this can be assessed to be around $2 \times 10^{-9}$, this shows that also the relative SGE precision of the assessment of gas density is very high (in the $10^{-9}$ range). Since $\rho_n$ takes a value of 44 mol/m³ under STP conditions (for N₂), henceforth referred to as $\rho_n^{STP}$, this implies that $\sigma_{\rho_{n,i}}^{SGE}$ can be estimated to be around $9 \times 10^{-8}$ mol/m³. This case is illustrated by the blue horizontal solid curve (left and lower axes) in Fig. 2a.

### 2. Multiple gas evacuation precision

As for refraction, the MGE precision, $\sigma_{\rho_{n,i}}^{MGE}$, depends partly on the same parameters as the SGE precision, but also on the fluctuations in the residual gas density, $\sigma_{\rho_{n,f}}$. Hence it (as well as its relative counterpart, $\bar{\sigma}_{\rho_{n,i}}^{MGE}$, given by $\sigma_{\rho_{n,i}}^{MGE} / \rho_{n,i}$) can be expressed as

$$\bar{\sigma}_{\rho_{n,i}}^{MGE} = \frac{\sigma_{\rho_{n,i}}^{MGE}}{\rho_{n,i}} = \sqrt{(\bar{\sigma}_{\rho_{n,i}}^{SGE})^2 + \bar{\sigma}_{\rho_{n,f}}^2}, \qquad (23)$$

where $\bar{\sigma}_{\rho_{n,f}}$ is the contribution to the relative precision in the assessment of gas density from the relative fluctuations in the residual gas density, which, by the use of the Eqs (3) and (4), can be written as

$$\bar{\sigma}_{\rho_{n,f}} = \frac{\sigma_{\rho_{n,f}}}{\rho_{n,i}}. \qquad (24)$$

As before, unless special precautions are taken, it is sensible to assume that the relative fluctuations in the residual gas density, $\bar{\sigma}_{\rho_{n,f}}$, are larger than the SGE precision, $\bar{\sigma}_{\rho_{n,i}}^{SGE}$. However, as was discussed above, a remedy to this is to monitor the pressure of the residual gas after each evacuation and use this to compensate the assessment of gas by use of Eq. (10) in the Eqs. (3) and (4). This can reduce $\bar{\sigma}_{\rho_{n,f}}$ to

$$\bar{\sigma}_{\rho_{n,f}} = \frac{1}{\rho_{n,i}}\frac{\sigma_{p_f}}{RT} = \frac{n_f - 1}{n_i - 1}\frac{3A_R}{2(n_f - 1)}\frac{\sigma_{p_f}}{RT} \approx \frac{p_f}{p_i}\frac{\sigma_{p_f}}{p_f}, \qquad (25)$$

where, $\sigma_{p_f}$, as above, represents the precision of repeated gas evacuations.

A comparison of the relative SGE precision, i.e. $\bar{\sigma}_{\rho_{n,i}}^{SGE}$, and the relative precision of the residual gas density, i.e. $\bar{\sigma}_{\rho_{n,f}}$, shows again that the former will dominate when the precision of the pressure gauge, $\sigma_p$, fulfills Eq. (12). When this is the case, $\bar{\sigma}_{\rho_{n,i}}^{MGE}$ is given by $\bar{\sigma}_{\Delta v_l}$, i.e. Eq. (21), which, in this work, has been assessed to $2 \times 10^{-9}$.

When the opposite holds, the relative MGE precision of the assessment of refractivity by FS-DFPC-OR, $\bar{\sigma}_{\rho_{n,i}}^{MGE}$, will be given by $\bar{\sigma}_{\rho_{n,f}}$, which, according to Eq. (25), at STP conditions, can be estimated to be $10^{-5}$ $\sigma_{p_f}$. Hence, for the case $\sigma_{p_f}$ is 1 mPa, $\bar{\sigma}_{\rho_{n,i}}^{MGE}$ becomes $10^{-8}$, whereby $\sigma_{\rho_{n,i}}^{MGE}$ is around $5 \times 10^{-7}$ mol/m³. For illustrative purposes, the case when the residual gas pressure is monitored by a commercially available pressure gauge with a maximum range of 0.1 Torr (13 Pa) and a $\sigma_{p_f}$ of 0.4 mPa (CTR 100 0.1, Leybold) is given by the blue horizontal dotted curve (left and lower axes) in Fig. 2a.

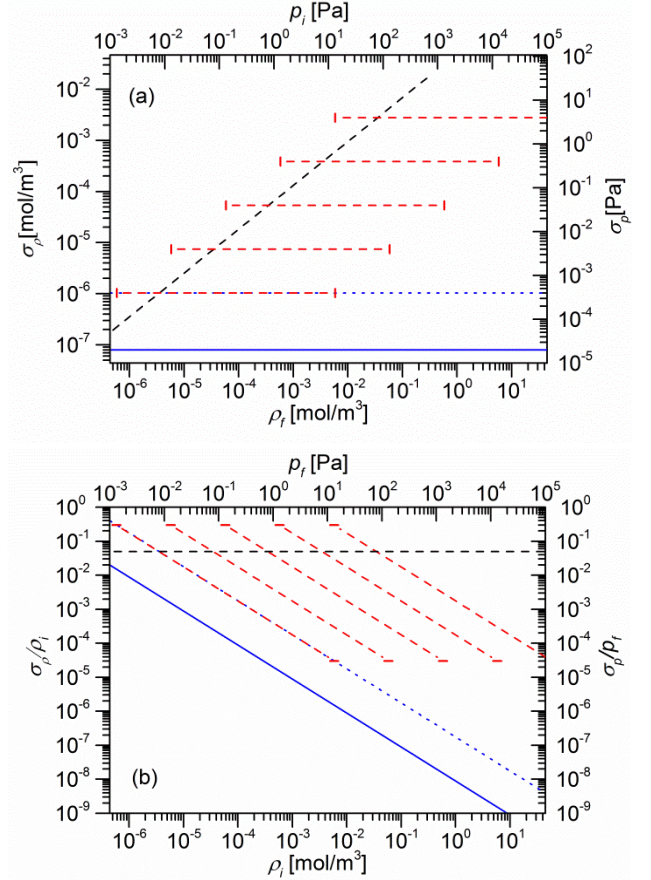

FIG 2. Panel (a) Blue curves, left y-axis and lower x-axis: schematic illustration of the precision of a gas density assessment as a function of the initial gas density by FS-DFPC-OR. The horizontal solid and dotted curves represent the precision when it is limited by the SGE precision, i.e. $\sigma_{\rho_{n,i}}^{SGE}$, and the residual gas density (assessed by a CTR 100 pressure gauge with a maximum range of 0.1 Torr), i.e. $\sigma_{\rho_{n,f}}$, respectively. Red and black dashed curves, right y-axis. and upper x-axis: schematic illustration of the precision of a gas pressure assessment as a function of the final gas pressure by a selection of conventional pressure gauges, i.e. various $\sigma_{p_f}$. The horizontal dashed curves represent the situations when the pressures are assessed by pressure gauges working in various ranges (from the bottom to the top: CTR 100 0.1/1/10/100/1000, where the second number represents the maximum pressure range in Torr), while the black dashed curve represents an ITR90 (all Leybold). Panel (b) Schematic illustration of the corresponding relative entities. Blue curves, left y-axis, and lower x-axis: the relative precision of a gas density assessment as a function of the initial gas density by FD-FPC-OR. The solid and dotted curves represent the relative precisions when the density assessment is limited by the relative SGE precision and the residual gas density [assessed by the same pressure gauge as in panel (a)], i.e. $\bar{\sigma}_{\rho_{n,i}}^{SGE}$ and $\bar{\sigma}_{\rho_{n,f}}$, respectively. Red and black dashed curves, right y-axis, and upper x-axis: schematic illustration of the relative precision of a gas pressure assessment as a function of the final gas pressure by the same set of pressure gauges as in panel (a), i.e. $\sigma_{p_f} / p_f$.

It is also of importance to notice that, irrespective of whether Eq. (12) is fulfilled or not, the precision of the assessment of gas density, $\sigma_{\rho_{n,i}}^{MGE}$, which is given by

$$\sigma_{\rho_{n,i}}^{MGE} = \sqrt{\left(\frac{2\tilde{\chi}_{\Delta q}}{3A_R}\frac{\sigma_{\Delta v_l}}{v_0}\right)^2 + \left(\frac{\sigma_{p_f}}{RT}\right)^2}, \qquad (26)$$

is independent of the gas density to be assessed, $\rho_{n,i}$. This is the reason why the two aforementioned blue curves in Fig. 2a are horizontal. This is in contrast to a majority of the alternative devices that can be used for assessment of gas density, e.g. pressure gauges, whose precision often are proportional to the range of the pressure gauge. This is schematically illustrated by the red horizontal dashed curves (right and upper axes) in Fig. 2a for a set of commercially available pressure gauges with maximum pressure readings that range from of 0.1 to 1000 Torr (13 to 13 × 10$^5$ Pa) that have a $\sigma_p$ of 0.003% of full scale, thus ranging from 0.4 mPa to 4 Pa (CTR 100 0.1/1/10/100/1000, Leybold). Note that in this figure the upper x-axis refers to the residual pressure, $p_f$, (which is in contrast to Fig. 1 in which it represents the pressure that corresponds to the gas density to be assessed, i.e. $p_i$). Alternatively, there are also combination devices that have larger dynamic ranges, which, according to manufactory specifications, have a resolution that is proportional to the pressure. One such (ITR90, Leybold) is exemplified by the black dashed curve (right and upper axes) in Fig. 2a. This figure shows, among other things, that it suffices to reduce the pressure to 13 Pa (0.1 Torr) to assess the residual gas density with a pressure gauge (the CTR 100 0.1) that provides a precision in the assessment of gas density of 10$^{-6}$ mol/m$^3$ (lowermost red curve), which, for assessment under STP conditions, corresponds to a relative precision of 2 × 10$^{-8}$.

All this implies, that the relative precision of assessments of gas density by FS-DFPC-OR, which often represents the relative resolution of an instrumentation, $\bar{\sigma}_{\rho_{n,i}}^{MGE}$, is inversely proportional to the gas density, while that of an assessment of pressure (or gas density by the use of a pressure gauge) is either so only for a limited range or pressures (shown by the red slanting curves in Fig. 2b) or, for a combination device, constant (displayed by the black horizontal curve in Fig. 2b). This demonstrates that the FS-DFPC-OR technique is particularly useful for assessment of gas densities under high density conditions since it can provide the highest relative precisions under such conditions.

**B. Accuracy of the assessment of gas density by FS-DFPC-OR**

*1. General description*

A scrutiny of Eq. (3) illustrates that the uncertainty of an assessment of gas density by FS-DFPC-OR, here denoted $\delta_{\rho_n}$, and its relative counterpart, $\bar{\delta}_{\rho_{n,i}}$, given by $\delta_{\rho_n} / \rho_{n,i}$, can, for the case with no relocking, be formally expressed as

$$\bar{\delta}_{\rho_{n,i}} = \frac{\delta_{\rho_{n,i}}}{\rho_{n,i}} = \sqrt{\bar{\delta}_{\Delta\nu_l}^2 + \bar{\delta}_{\nu_0}^2 + \bar{\delta}_{\tilde{\chi}_{\Delta\nu}}^2 + \bar{\delta}_{A_R}^2 + \bar{\delta}_{\tilde{B}_{\Delta\nu}}^2 + \bar{\delta}_{\tilde{C}_{\Delta\nu}}^2 + \bar{\delta}_{\rho_{n,f}}^2} \ , \quad (27)$$

where the $\bar{\delta}_{\tilde{\chi}_{\Delta\nu}}$, $\bar{\delta}_{A_R}$, $\bar{\delta}_{\tilde{B}_{\Delta\nu}}$, $\bar{\delta}_{\tilde{C}_{\Delta\nu}}$, and $\bar{\delta}_{\rho_{n,f}}$ are the individual contributions to the relative uncertainty of the assessment of gas density due to the uncertainties in $\tilde{\chi}_{\Delta\nu}$, $A_R$, $\tilde{B}_{\Delta\nu}$, $\tilde{C}_{\Delta\nu}$, and $\rho_{n,f}$, which are denoted $\delta_{\tilde{\chi}_{\Delta\nu}}$, $\delta_{A_R}$, $\delta_{\tilde{B}_{\Delta\nu}}$, $\delta_{\tilde{C}_{\Delta\nu}}$, and $\delta_{\rho_{n,f}}$, respectively. The various individual contributions are given by

$$\bar{\delta}_{\tilde{\chi}_{\Delta\nu}} = \frac{\delta_{\tilde{\chi}_{\Delta\nu}}}{\tilde{\chi}_{\Delta\nu}} \approx \delta_{\tilde{\chi}_{\Delta\nu}} \quad (28)$$

$$\bar{\delta}_{A_R} = \frac{\delta_{A_R}}{A_R} , \quad (29)$$

$$\bar{\delta}_{\tilde{B}_{\Delta\nu}} = \rho_{n,i} \frac{3 A_R}{2} \delta_{\tilde{B}_{\Delta\nu}} = (n_i - 1) \delta_{\tilde{B}_{\Delta\nu}} , \quad (30)$$

$$\bar{\delta}_{\tilde{C}_{\Delta\nu}} = \rho_{n,i}^2 \left(\frac{3 A_R}{2}\right)^2 \delta_{\tilde{C}_{\Delta\nu}} = (n_i - 1)^2 \delta_{\tilde{C}_{\Delta\nu}} , \quad (31)$$

and

$$\bar{\delta}_{\rho_{n,f}} = \frac{\delta_{\rho_{n,f}}}{\rho_{n,i}} , \quad (32)$$

respectively. Estimated magnitudes of these uncertainties for a typical well-designed but uncharacterized FS-DFCB-OR system are presented in Table 2 and Table 3. The data given in Table 3 are taken from the literature, where the various uncertainties have been estimated as follows.[20]

Table 3. Uncertainties in various entities for a typical uncharacterized FS-DFCB-OR instrumentation for assessment of gas density that are not listed in Table 2. Data taken from the literature

| Entity | | Value |
|---|---|---|
| Uncertainty of $\tilde{\chi}_{\Delta\nu}$ | $\bar{\delta}_{\tilde{\chi}_{\Delta\nu}}$ | 2 × 10$^{-5}$/2 × 10$^{-4 a,b}$ |
| Uncertainty of $A_R$ | $\bar{\delta}_{A_R}$ | 6 × 10$^{-4 b,c}$ |
| Uncertainty of $\tilde{B}_{\Delta\nu}$ | $\bar{\delta}_{\tilde{B}_{\Delta\nu}}$ | 3 × 10$^{-5 b,d}$ |
| Uncertainty of $\tilde{C}_{\Delta\nu}$ | $\bar{\delta}_{\tilde{C}_{\Delta\nu}}$ | 3 × 10$^{-8 b,e}$ |
| Uncertainty of the residual density assessment | $\bar{\delta}_{\rho_{n,f}}$ | 2 × 10$^{-7 f}$ |

[a] Taken as $\delta_\varepsilon$ as given in Table 2, i.e. first value: for a well-constructed closed cavity, second value: for an open cavity.
[b] Can be improved by a characterization procedure.
[c] For nitrogen. Estimated to be 2 × 10$^{-4}$ at 633 nm,[21] additional uncertainty at a wavelength of 1.5 μm estimated from is wavelength dependence.
[d] For nitrogen at STP conditions, assuming $\delta_{\tilde{B}_{\Delta\nu}} \sim 0.1$.
[e] For nitrogen at STP conditions, assuming $\delta_{\tilde{C}_{\Delta\nu}} \sim 0.1$.
[f] For STP condition, assuming an uncertainty in $p_f$ of 20 mPa [e.g. Ceravac, CTR 100, 1 Torr, Oerlicon], taken as $\delta_{n_f - 1}/(n_i - 1)$ from Table 2.

The uncertainty in $\tilde{\chi}_{\Delta\nu}$, i.e. $\delta_{\tilde{\chi}_{\Delta\nu}}$, is formally given by $[\delta_\varepsilon^2 + (2\delta_\varsigma)^2 + \delta_\eta^2]^{1/2}$. However, since $\delta_\varepsilon$ is significantly larger than the other entities, $\bar{\delta}_{\tilde{\chi}_{\Delta\nu}}$ is, in practice, given by $\bar{\delta}_\varepsilon$, which, according to Table 2, can be estimated to 2 × 10$^{-5}$ or 2 × 10$^{-4}$ depending on type of cavity that is used. The relative uncertainty of the $A_R$ coefficient at the wavelength considered (i.e. 1.5 μm) was estimated from the uncertainty of the same coefficient at 633 nm, which was assessed by Achterman et al.[21] to 2 × 10$^{-4}$. It was then cautiously assumed that the wavelength dependence of the $A_R$ coefficient assessed by Peck et al.[22] will provide a relative uncertainty to the assessment of gas density by FS-DFPC-OR that is twice this value. The overall uncertainty of the $A_R$ coefficient at the wavelength around 1.5 μm has therefore been taken as 6 × 10$^{-4}$. The uncertainties in the $\tilde{B}_{\Delta\nu}$ and $\tilde{C}_{\Delta\nu}$ coefficients, i.e. the $\delta_{\tilde{B}_{\Delta\nu}}$ and $\delta_{\tilde{C}_{\Delta\nu}}$, whose leading terms are $(5/6)[1 - (4/5)(B_R/A_R^2)]$ and $(4/9)[1 - (5/2)(B_R/A_R^2) - (C_R/A_R^3)]$, respectively, have been taken as 0.1 and 0.3, respectively. Since, under STP conditions, $n_i - 1$ typically takes a value of 3 × 10$^{-4}$, this implies that the contributions to $\bar{\delta}_{\rho_{n,i}}$ due to the uncertainties in these two, i.e. $\bar{\delta}_{\tilde{B}_{\Delta\nu}}$ and $\bar{\delta}_{\tilde{C}_{\Delta\nu}}$, become 3 × 10$^{-5}$ and 3 × 10$^{-8}$, respectively. Finally, the uncertainty of the residual density assessment has been taken as that measured by a pressure gauge with an accuracy of 0.2% of the reading at a pressure of 10 Pa. This implies that the contributions to the relative uncertainty of the assessment of gas density under STP conditions due to this, $\bar{\delta}_{\rho_{n,f}}$, becomes 2 × 10$^{-7}$.

*2. Accuracy of an uncharacterized system*

Based on Table 3, Fig. 3a displays the individual contributions to the relative uncertainty of the gas density, $\bar{\delta}_{\rho_{n,i}}$, as a function of gas density (lower x- axis) as well as pressure (upper x- axis) for a well-designed FS-DFCB-OR system run without relocking that is not characterized with respect to a known density. For illustrational purposes, the individual contributions to the relative uncertainty of the gas density are schematically displayed as a function of gas density (and pressure) in Fig. 3a and under STP conditions in Fig. 3b.

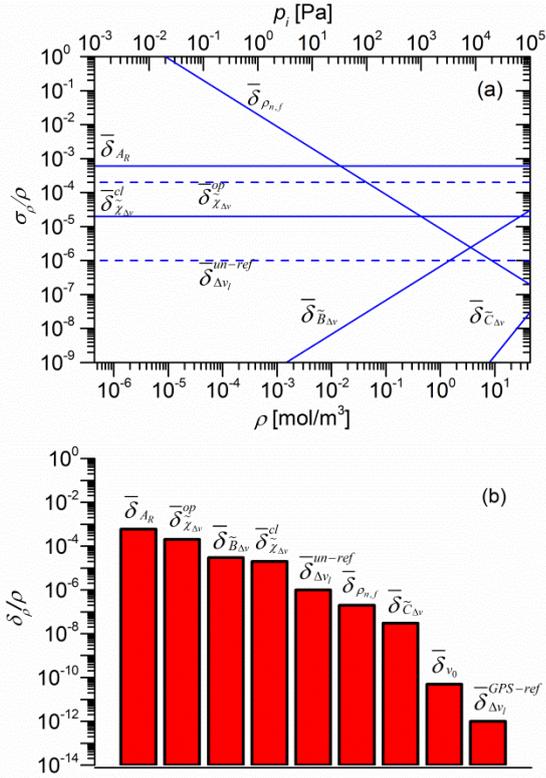

FIG 3. Panel (a) Schematic illustration of the relative contributions to the uncertainty of the gas density, $\bar{\delta}_{\rho_{n,i}}$, expressed as $\delta_{\rho_n}/\rho_{n,i}$, from various physical entities in terms of the gas density to be assessed, $\rho_{n,i}$, (lower x-axis), and, for clarity, the corresponding pressure, $p_i$, (upper x-axis). The horizontal curves represent, from top to bottom, the relative contributions from the $A_R$ coefficient, $\bar{\delta}_{A_R}$, the $\tilde{\chi}_{\Delta\nu}$ coefficient, $\bar{\delta}_{\tilde{\chi}_{\Delta\nu}}$, and the shift of the laser frequency for a unreferenced mode of operation of the frequency counter, $\bar{\delta}_{\Delta\nu_l}^{un-ref}$. The uppermost slanted solid curve (decreasing with density) corresponds to the contribution from the residual refractivity, $\bar{\delta}_{\rho_{n,f}}$, while the two lowermost slanting curves (increasing with density) represent the contribution from the virial coefficients, $\bar{\delta}_{\tilde{B}_{\Delta\nu}}$ and $\bar{\delta}_{\tilde{C}_{\Delta\nu}}$, respectively. Panel (b) Schematic illustration of the corresponding situation under STP conditions, including also the contributions from the laser frequency, $\bar{\delta}_{\nu_0}$, and the shift of the laser frequency for the case with a frequency counter that is referenced to a GPS, $\bar{\delta}_{\Delta\nu_l}^{GPS-ref}$. In both panels, the $\bar{\delta}_{\tilde{\chi}_{\Delta\nu}}^{cl}$ and $\bar{\delta}_{\tilde{\chi}_{\Delta\nu}}^{op}$ refer to the cases with a well-constructed closed cavity and an open cavity, respectively.

Figure 3a shows that also for the case with the relative uncertainty of the gas density, the contribution from the relative uncertainty of the residual refractivity, $\bar{\delta}_{\rho_{n,f}}$, decreases with gas density (pressure). Although there again are some entities whose contributions to $\bar{\delta}_{\rho_{n,i}}$ is independent of density (pressure), $\bar{\delta}_{A_R}$, $\bar{\delta}_{\tilde{\chi}_{\Delta\nu}}$, and $\bar{\delta}_{\Delta\nu_l}^{un-ref}$, there are also two whose contributions increase with density, viz. the virial coefficients, $\bar{\delta}_{\tilde{B}_{\Delta\nu}}$ and $\bar{\delta}_{\tilde{C}_{\Delta\nu}}$. Among the ones that are independent of density, it is possible to conclude that, under the pertinent conditions, and for the case with a closed cavity, the relative uncertainty of $A_R$, i.e. $\bar{\delta}_{A_R}$, will dominate over the other two. If an open cavity is used, though, it is possible that the uncertainty of the $\tilde{\chi}_{\Delta\nu}$ coefficient, i.e. $\bar{\delta}_{\tilde{\chi}_{\Delta\nu}}$, can start to rival that of the $A_R$ coefficient, i.e. $\bar{\delta}_{A_R}$.

For sufficiently high densities, in this particular case above 0.02 mol/m$^3$ (∼ 40 Pa), the contribution from $\bar{\delta}_{A_R}$ will dominate also over that from the residual refractivity. At low densities, on the other hand, the uncertainty of a gas density assessment is dominated by the uncertainty from the residual refractivity.

For the situation considered, and for a closed cavity under STP conditions, with the present uncertainty of $A_R$, none of the other entities will contribute significantly to the uncertainty in the density.

Since the leading term of the expression for the gas density for the case with relocking, i.e. Eq. (4), has the same linear dependence on $A_R$ as that with no relocking, given by Eq. (3), a system incorporating relocking will have an accuracy that is similar to (in practice equal to) that of a system with no relocking.

This implies that an uncharacterized FS-DFCB-OR system cannot be expected to provide a relative accuracy for assessment of gas density that is better than that of the $A_R$ coefficient, i.e. $\bar{\delta}_{A_R}$, which presently is in the mid-10$^{-4}$ range. However, as is discussed in more detail in section VI, if the system can be characterized by one type of gas whose density is well-known (i.e. known with a low uncertainty) under at least one set of conditions, the system will demonstrate a significantly better accuracy.

## V. TEMPERATURE DEPENDENCE OF THE ASSESSMENT OF REFRACTIVITY AND GAS DENSITY BY FAST SWITCHING FABRY-PEROT-BASED OPTICAL REFRACTOMETRY

Since many processes have a temperature dependence, unless meticulously designed, every instrumentation will be affected by a change in temperature. This is also the case with OS. However, with knowledge about how such processes affect measurements, means to alleviate their influence can often be taken, resulting in instrumentation and procedures that demonstrate a minimal dependence on temperature.

The major contributions to a DFPB-OR signal from a drifting temperature originate from the thermal expansion of the spacer material and the alteration of the instantaneous change in length of the cavity due to the presence of the gas in the cavity. As was alluded to above, FPC-OR methodologies are based on a measurement of the shift of the frequency of the laser locked to the measurement cavity as the cavity, which originally is filled with gas, is evacuated. The basis for the DF-(D)FPC-OR methodologies scrutinized in this as well as our accompanying papers[13, 14] is that the evacuation is made so swift that "slow" processes, primarily temperature drifts and relaxations, do not influence the assessment, so as to make the assessment "drift free". It is prophesized that this can be realized by making "fast switching" of the gas in the measurement cavity, performed on a time scale that is faster than the influence of any substantial drift or relaxation, in practice solely for the time period under which the cavity is evacuated.[13] This implies, among other things, that the technique is differently sensitive to temperature fluctuations/drifts that take place *during* the time under which the gas is evacuated (i.e. when an actual refractivity measurement is performed) and those that appear *between* such measurements. These two situations should therefore preferably be considered separately.

### A. Influence of temperature drift on the shift of the laser frequency

*1. Effect of drift of the temperature between evacuations*

As is shown in the supplementary material, it is possible to derive an expression for the change in the shift of the cavity mode frequency (and thereby, for the case with no relocking, the change in the shift of the laser beat frequency) that takes place while the cavity is being evacuated, $\delta_{\Delta\nu_l}(\delta T_B)$, when the temperature drifts *between* evacuations of the cavity an amount $\delta T_B$, from a temperature $T_0$ to a temperature of $T_0 + \delta T_B$. The Eqs. (S1) and (S2) show that for the case when we can consider the index of refraction of the gas to be independent of the temperature, i.e. that $(n_i)_{T_0} = (n_i)_{T_0+\delta T_B}$, this entity, and its relative counterpart, denoted $\bar{\delta}_{\Delta\nu_l}(\delta T_B)$ and defined as $\delta_{\Delta\nu_l}(\delta T_B)/\Delta\nu_l$, can be expressed as

$$\bar{\delta}_{\Delta v_l}(\delta T_B) = \frac{\delta_{\Delta v_l}(\delta T_B)}{\Delta v_l}$$

$$= \frac{1}{\Delta v_l} \left( \left\{ [v_q(n_f)] - [v_q(n_i)] \right\}_{T_0 + \delta T_B} \right.$$

$$\left. - \left\{ [v_q(n_f)] - [v_q(n_i)] \right\}_{T_0} \right) \quad (33)$$

$$\approx \left[ \frac{\varepsilon(T_0)}{T_0} - \alpha \right] \delta T_B,$$

where $\varepsilon(T_0)$ is the parameter that describe the instantaneous influence of the gas on the length of the cavity and $\alpha$ is the thermal expansion coefficient of the spacer material.

*2. Effect of drift of the temperature during an evacuation*

The supplementary material also derives, by the Eqs. (S3) and (S4), expressions for the change in the shift of the beat frequency that takes place while the cavity is being evacuated that originates from a temperature drift of $\delta T_D$ that takes place *during* the evacuation, from a temperature $T$ to a temperature of $T + \delta T_D$, denoted $\delta_{\Delta v_l}(\delta T_D)$, and its relative counterpart, denoted $\bar{\delta}_{\Delta v_l}(\delta T_D)$ and defined as $\delta_{\Delta v_l}(\delta T_D) / \Delta v_l$, where again the latter can be expressed as

$$\bar{\delta}_{\Delta v_l}(\delta T_D) = \frac{\delta_{\Delta v_l}(\delta T_D)}{\Delta v_l}$$

$$= \frac{1}{\Delta v_l} \left( \left\{ [v_q(n_f)]_{T + \delta T_D} - [v_q(n_i)]_T \right\} \right.$$

$$\left. - \left\{ [v_q(n_f)]_T - [v_q(n_i)]_T \right\} \right) \quad (34)$$

$$= -\frac{v_0(T)}{\Delta v_l} \left[ \alpha + (n_f - 1) \frac{\varepsilon(T)}{T} \right] \delta T_D$$

$$= -\left[ \frac{\alpha}{n_i - 1} + \frac{n_f - 1}{n_i - 1} \frac{\varepsilon(T)}{T} \right] \delta T_D \approx -\frac{\alpha}{n_i - 1} \delta T_D,$$

where $v_0(T)$ is the frequency of the empty cavity mode addressed at a temperature of $T$. Note that we in the last line have made use of the fact that the methodology assumes that the cell is evacuated to such an extent that $(n_f - 1)/(n_i - 1) \ll 1$, which implies that the $\varepsilon$-term can be neglected with respect to the $\alpha$-term.

*3. Comparison between drift of the temperature between and during evacuations*

It can first of all be concluded that although both types of changes in $\Delta v_l$, i.e. the two $\delta_{\Delta v_l}$, depend on the thermal expansion of the spacer material (i.e. on $\alpha$), the effect of the thermal expansion of the spacer material is significantly smaller (several orders of magnitude) when the drift takes place *between* evacuations than *during*; while $\alpha$ is in the order of $10^{-8}$ or $10^{-7}$ K$^{-1}$ under some general STP conditions, $\alpha/(n_i - 1)$ can typically range from $3 \times 10^{-5}$ to $3 \times 10^{-4}$ K$^{-1}$, where in both cases, the lower value represents the case with a spacer made by a ULE material. This shows that the shift of the beat frequency due to thermal expansion of the spacer material is typically $3 \times 10^3$ times smaller for drifts that take place between evacuations than during.

In addition, the expressions above also show that the shift of the beat frequency has a dependence on temperature originating from the instantaneous change in length of the cavity, $\varepsilon$, for the case when the temperature drift takes place between the evacuations [Eq. (33)] but, in practice, not during [Eq. (34)]. The reason is that a change in temperature affects the instantaneous change in length of the cavity in proportion to the gas pressure. Any change in temperature during the evacuation will then not affect the shift of the beat frequency since there is a low gas pressure in the final state; it is only the temperature in the initial state that can affect the beat frequency.

Moreover, since, under STP conditions, $\varepsilon$ can have a value in the low $10^{-4}$ range for a well-constructed closed cavity while it can take a value in the low $10^{-3}$ range for an open one, the $\varepsilon(T_0)/T_0$ term will take values in the low $10^{-6}$ to the low $10^{-5}$ K$^{-1}$ range. Since $\alpha$ is often in the order of $10^{-8}$ or $10^{-7}$ K$^{-1}$, this implies that the relative change of the shift of the beat frequency due to a change in the temperature between evacuations, given by Eq. (33), is in practice dominated by the temperature dependence of the instantaneous change in length of the cavity from the presence of the gas, given by the $\varepsilon(T_0)/T_0$-term. Hence, $\bar{\delta}_{\Delta v_l}(\delta T_B)$ can be considered to be mainly given by

$$\bar{\delta}_{\Delta v_l}(\delta T_B) = \frac{\delta_{\Delta v_l}(\delta T_B)}{\Delta v_l} \approx \frac{\varepsilon(T_0)}{T_0} \delta T_B. \quad (35)$$

This implies that the total temperature dependence of the relative change of the shift of the beat signal, $\bar{\delta}_{\Delta v_l}(\delta T_D, \delta T_B)$, can, in practice, be expressed as

$$\bar{\delta}_{\Delta v_l}(\delta T_B, \delta T_D) = \bar{\delta}_{\Delta v_l}(\delta T_B) + \bar{\delta}_{\Delta v_l}(\delta T_D)$$

$$= \frac{\varepsilon(T_0)}{T_0} \delta T_B - \frac{\alpha}{n_i - 1} \delta T_D. \quad (36)$$

Moreover, since $\varepsilon(T_0)/T_0$ will be in the low $10^{-6}$ to the low $10^{-5}$ K$^{-1}$ range while $\alpha/(n_i - 1)$ typically ranges from $3 \times 10^{-5}$ to $3 \times 10^{-4}$ K$^{-1}$ under some general STP conditions, this shows that *the temperature dependence of the shift of the beat frequency is **significantly smaller** for drifts in temperature **between evacuations** than **during***; about a factor of three for the case with an open cavity made of ULE and about a factor of 300 for a closed cavity made of Zerodur.[23]

Since the change in the shift of the beat frequency due to a drift in temperature, $\delta_{\Delta v_l}(\delta T)$, is related to its relative counterpart as $(n_i - 1)\bar{\delta}_{\Delta v_l}(\delta T)v_0$, this implies that it can be expressed as

$$\delta_{\Delta v_l}(\delta T_B, \delta T_D) = v_0 \left[ \frac{\varepsilon(T_0)}{T_0}(n_i - 1)\delta T_B - \alpha \delta T_D \right]. \quad (37)$$

This shows that the change in the shift of the beat frequency due to a drift in temperature between evacuations, $\delta T_B$, is proportional to the refractivity while that during evacuations, $\delta T_D$, is not. This indicates that a drift *between* evacuations can be seen as a *reduction of the accuracy* of the technique while a drift *during* evacuations can be seen as a *reduction of the precision* (or the resolution) of the technique.

**B. Influence of temperature drift on the assessment of refractivity**

When estimating the influence of a temperature drift during evacuation of the cavity on the assessment of refractivity it suffices to look at the influence of the temperature drift on the leading term of the expression for the entity considered. For the case when refractivity is assessed, and for the case with no relocking, we will therefore assume that the refractivity is given by the leading terms of Eq. (1), i.e., when measured as a temperature of $T$, by

$$(n_i - 1)_T = \Omega(\varepsilon, \varsigma, \eta) \frac{\Delta v_l}{v_0} = [1 - \varepsilon(T)] \frac{\Delta v_l(T)}{v_0(T)}. \quad (38)$$

Let us moreover assume that the system is characterized at a given temperature, $T_0$, so we know that at this temperature, a given shift in beat frequency, $\Delta v_l(T_0)$, measured for a given frequency of the cavity mode addressed in an empty cavity, i.e. $v_0(T_0)$, corresponds to a given refractivity, here denoted $(n_i - 1)_{T_0}$. This implies that we can assume that the expression

$$(n_i - 1)_{T_0} = [1 - \varepsilon(T_0)] \frac{\Delta v_l(T_0)}{v_0(T_0)} \quad (39)$$

is valid. This can be seen as a characterization of the system, in this case the $\varepsilon(T_0)$ - parameter.

Consider then the situation when the instrumentation is used for assessment of the same refractivity, i.e. $(n_i - 1)_{T_0}$, but under conditions, unknown to the user, that the temperature has drifted an amount of $\delta T_B$ since the last characterization, i.e. from $T_0$ to $T = T_0 + \delta T_B$, and drifts $\delta T_D$ during the measurement, i.e. from $T$ to $T + \delta T_D$. Such a measurement will then provide a shift of the beat frequency, $\Delta \nu_l(T_0, \delta T_B, \delta T_D)$, and the frequency of the cavity mode addressed by the laser light in an empty cavity, $\nu_0(T_0, \delta T_B, \delta T_D)$, that are given by

$$\Delta \nu_l(T_0, \delta T_B, \delta T_D) = \Delta \nu_l(T_0)\left[1 + \bar{\delta}_{\Delta \nu_l}(\delta T_B, \delta T_D)\right] \quad (40)$$

and

$$\nu_0(T_0, \delta T_B, \delta T_D) = \nu_0(T_0)\left[1 + \bar{\delta}_{\nu_0}(\delta T_B, \delta T_D)\right], \quad (41)$$

where $\bar{\delta}_{\Delta \nu_l}(\delta T_B, \delta T_D)$ is the relative change in the shift of the beat frequency due to the changes in temperature, given by Eq. (36), and $\bar{\delta}_{\nu_0}(\delta T_B, \delta T_D)$ is the relative change in the frequency of the cavity mode addressed by the laser light in an empty cavity, given by Eq. (S6) in the supplementary material.

Such an assessment will then provide an apparent refractivity value, $(n_i - 1)_{T_0, \delta T_D, \delta T_B}$, that will differ from the actual value, $(n_i - 1)_{T_0}$, by a relative amount $\bar{\delta}_{n_i - 1}(\delta T_B, \delta T_D)$, given by

$$(n_i - 1)_{T_0, \delta T_D, \delta T_B} = (n_i - 1)_{T_0}\left[1 + \bar{\delta}_{n_i - 1}(\delta T_B, \delta T_D)\right]. \quad (42)$$

This implies that, by the use of Eq. (38), the relative error in the assessment of the refractivity value can be related to the shifts of the beat frequency and the frequency of the cavity mode addressed by the laser light in an empty cavity according to

$$(n_i - 1)_{T_0}\left[1 + \bar{\delta}_{n_i - 1}(\delta T_B, \delta T_D)\right] = \\ \left[1 - \varepsilon(T_0)\right]\frac{\Delta \nu_l(T_0)\left[1 + \bar{\delta}_{\Delta \nu_l}(\delta T_B, \delta T_D)\right]}{\nu_0(T_0)\left[1 + \bar{\delta}_{\nu_0}(\delta T_B, \delta T_D)\right]}, \quad (43)$$

where we have assumed that the user, unaware of the change in temperature, relies on the $\varepsilon$ -parameter obtained by the calibration, i.e. $\varepsilon(T_0)$.

Since all relative entities can be assumed to be small, i.e. $\ll 1$, this expression shows, together with the characterization condition, i.e. Eq. (39), that the relative error in the assessment of the refractivity, $\bar{\delta}_{n_i - 1}(\delta T_B, \delta T_D)$, is given by[24]

$$\bar{\delta}_{n_i - 1}(\delta T_B, \delta T_D) = \bar{\delta}_{\Delta \nu_l}(\delta T_B, \delta T_D) - \bar{\delta}_{\nu_0}(\delta T_B, \delta T_D). \quad (44)$$

Inserting the explicit expressions for the $\bar{\delta}_{\Delta \nu_l}(\delta T_B, \delta T_D)$ entity from Eq. (36) and $\bar{\delta}_{\nu_0}(\delta T_B, \delta T_D)$ from Eq. (S6) in the supplementary material shows that $\bar{\delta}_{n_i - 1}(\delta T_B, \delta T_D)$ is given by

$$\bar{\delta}_{n_i - 1}(\delta T_B, \delta T_D) = \frac{\varepsilon(T_0)}{T_0}\delta T_B - \frac{\alpha}{n_i - 1}\delta T_D + \alpha(\delta T_B + \delta T_D) \\ = \left(\frac{\varepsilon(T_0)}{T_0} + \alpha\right)\delta T_B - \left(\frac{1}{n_i - 1} - 1\right)\alpha \delta T_D. \quad (45)$$

Since $\varepsilon(T_0)/T_0 > \alpha$, the last term in the first set of parentheses can be neglected with respect to the first. In addition, since $(n_i - 1)^{-1}$ is significantly larger than unity, also the last term in the second set of parentheses can be neglected with respect to the first.[25] This implies that the relative error in the assessment of the refractivity, $\bar{\delta}_{n_i - 1}(\delta T_B, \delta T_D)$, can, in practice, be considered to be given by

$$\bar{\delta}_{n_i - 1}(\delta T_B, \delta T_D) = \frac{\varepsilon(T_0)}{T_0}\delta T_B - \frac{\alpha}{n_i - 1}\delta T_D. \quad (46)$$

It can be noticed that this expression is identical to Eq. (36). This implies that much of the conclusions drawn for the relative shift of the beat signal are applicable also to the relative error in the assessment of the refractivity. Hence, also the relative error in the assessment of the refractivity is significantly smaller for the case with drifts in temperature between evacuations ($\delta T_B$) than during ($\delta T_D$); about a factor of three for the case with an open cavity made of ULE and about a factor of 300 for a closed cavity made of Zerodur.

Since, for a well-designed closed cavity, $\varepsilon(T_0)/T_0$ will take values in the low $10^{-6}$ K$^{-1}$ region, if the drifts in the temperature between evacuations can be kept at a level of 1 mK, Eq. (46) shows that this drift will solely contribute to a relative error in the assessment of the refractivity of $10^{-9}$, which corresponds to an absolute error in refractivity of $3 \times 10^{-13}$.

Moreover, since for ULE glass and under STP conditions, $\alpha/(n_i - 1)$ is in the order of $3 \times 10^{-5}$ K$^{-1}$, Eq. (46) also shows that, for the case the drifts in the temperature during the evacuation can be kept at the same level (i.e. 1 mK), the relative error in the assessment of the refractivity becomes $3 \times 10^{-8}$. This corresponds, under STP conditions, to an absolute error in refractivity of $10^{-11}$.

It can also be concluded that as long as both $\Delta \nu_l(T)$ and $\nu_0(T)$ are monitored and the mode number is known, long term drifts of the system (and thereby relative uncertainties in the assessment of refractivity) are solely given by $\varepsilon(T_0)/T_0$, which above was estimated to be in the low $10^{-6}$ K$^{-1}$ to the low $10^{-5}$ K$^{-1}$ range, which is significantly lower than what is obtained if the fast switching methodology would not be employed.

This shows that the FS-DFCB-OR technique has an exceptional small dependence on temperature when applied to assessments of refractivity and that it therefore is particularly suitable for applications under conditions when good temperature stability cannot be guaranteed.

For the case with relocking, the situation is similar to that without.

## C. Influence of temperature drift on the assessment of gas density

When gas density is to be assessed, and again for the case with no relocking, we will assume that it is given by the leading terms of Eq. (3), i.e., when measured at a temperature of $T$, by[26]

$$(\rho_{n,i})_T = \frac{2}{3A_R(T)}\tilde{\chi}_{\Delta \nu}(T)\frac{\Delta \nu_l(T)}{\nu_0(T)}\left\{1 + \tilde{B}_{\Delta \nu}(T)\frac{\Delta \nu_l(T)}{\nu_0(T)}\right\}, \quad (47)$$

where $\tilde{\chi}_{\Delta \nu}(T)$ is considered to be equal to $\Omega(\varepsilon, \varsigma, \eta, T)$, which we here take as its leading order, i.e. as $[1 - \varepsilon(T)]$.

Let us again assume that the system is characterized at a given temperature, $T_0$, so we know that at this temperature, a given shift in beat frequency, $\Delta \nu_l(T_0)$, measured for a given frequency of the cavity mode addressed in an empty cavity, i.e. $\nu_0(T_0)$, corresponds to a given gas density, here denoted $(\rho_{n,i})_{T_0}$. This implies that we can assume that the expression

$$(\rho_{n,i})_{T_0} = \frac{2}{3A_R(T_0)}\tilde{\chi}_{\Delta \nu}(T_0)\frac{\Delta \nu_l(T_0)}{\nu_0(T_0)}\left\{1 + \tilde{B}_{\Delta \nu}(T_0)\frac{\Delta \nu_l(T_0)}{\nu_0(T_0)}\right\} \quad (48)$$

is valid. This can be seen as a characterization of the system, primarily the $A_R(T_0)$, $\tilde{\chi}_{\Delta \nu}(T_0)$, and $\tilde{B}_{\Delta \nu}(T_0)$ - parameters.

Consider then the situation when the instrumentation is used for assessment of the same gas density as by which it was characterized, i.e. $(\rho_{n,i})_{T_0}$, for the case when the temperature has drifted an amount of $\delta T_B$ since the last characterization, from $T_0$ to $T = T_0 + \delta T_B$, and that it drifts $\delta T_D$ during an evacuation, from $T$ to $T + \delta T_D$, unwittingly to the user. Such a measurement will then provide a shift of the beat frequency, $\Delta \nu_l(T_0, \delta T_B, \delta T_D)$, and the frequency of the cavity mode addressed, $\nu_0(T_0, \delta T_B, \delta T_D)$, according to the Eqs. (40) and (41).

However, when we consider the influence of drifts of the temperature on the assessment of a given gas density, we cannot assume that $\bar{\delta}_{\Delta \nu_l}(\delta T_D, \delta T_B)$ can be given by Eq. (36) as was done when the influence of drifts in the temperature on the assessment of a given refractivity was considered. The reason for this is that Eq. (36) was derived under the assumption that the refractivity in the cavity is independent of the temperature; i.e. by tacitly assuming that $(n_i)_{T_0 + \delta T_B + \delta T_D}$ is identical to $(n_i)_{T_0}$. It was shown by Eq. (S48) in our accompanying paper[14] that for a given density, the refractivity of a gas depends on the temperature through the higher order virial coefficients, viz. as $n - 1 = A_m \rho_n + B_m \rho_n^2 + ...$, where $B_m$ is given by $3A_R^2/8 + 3B_R/2$ where, in turn, $A_R$ and $B_R$ are the first two virial coefficients in the extended Lorentz-Lorenz equation [Eq. (15) in the same reference]. Although $A_R$ is considered independent of temperature (it depends solely on the dipolar polarizability and the diamagnetic susceptibility of the gas), $B_R$ has a pronounced temperature dependence.[2] Hence, for a given density, $(n_i - 1)_{T_0 + \delta T_B + \delta T_D}$ will be dissimilar from $(n_i - 1)_{T_0}$.

It is shown in the supplementary material that the effect of a temperature dependent $B_R$ virial coefficient is to provide the refractivity with a temperature dependence, $(n - 1)(T)$, according to

$$(n-1)(T) = (n-1)(T_0) \cdot \left\{ 1 + \bar{\delta}[(n-1), \delta T_B] \right\}, \quad (49)$$

where $\bar{\delta}[(n-1), \delta T_B]$ is the relative change in the refractivity due to the change of the temperature that, under the assumption that $B_R(T)$ has a linear dependence on molecular velocity, is given by [see. Eq. (S10)]

$$\bar{\delta}[(n-1)(T_0), \delta T_B] = \frac{1}{3T_0} \frac{B_R(T_0)}{A_R^2} [(n-1)(T_0)] \delta T_B. \quad (50)$$

Since $B_R(T_0)/A_R^2$ is around 0.032 for the case considered, this implies, under STP conditions and according to Eq. (S11), that

$$\bar{\delta}[(n-1)(T_0), \delta T_B] = 1 \times 10^{-8} \cdot \delta T_B. \quad (51)$$

As is shown by Eq. (S12) in the supplementary material, this implies that the effect of a temperature dependent $B_R$ virial coefficient on the index of refraction, i.e. $n(T)$, can be written as

$$n(T) = n(T_0) \left\{ 1 + \bar{\delta}[n(T_0), \delta T_B] \right\}, \quad (51)$$

where $\bar{\delta}[n(T_0), \delta T_B]$ is the relative change in the index of refraction due to the change of the temperature, this is given by

$$\bar{\delta}[n(T_0), \delta T_B] = \frac{1}{3T_0} \frac{B_R(T)}{A_R^2} \frac{[(n-1)(T_0)]^2}{n(T_0)} \delta T_B. \quad (51)$$

As also is shown in the supplementary material, this implies that the relative change in the shift in the beat frequency, $\bar{\delta}_{\Delta \nu_l}(\delta T_B)$, no longer is given by Eq. (33), but rather by

$$\bar{\delta}_{\Delta \nu_l}(\delta T_B) = \frac{\delta_{\Delta \nu_l}(\delta T_B)}{\Delta \nu_l}$$
$$= \left[ \frac{1}{3T_0} \frac{B_R(T_0)}{A_R^2} (n_i - 1)(T_0) + \frac{\varepsilon(T_0)}{T_0} - \alpha \right] \delta T_B. \quad (51)$$

From this, it follows that the relative change in the shift of the beat frequency due to the changes in temperature, i.e. $\bar{\delta}_{\Delta \nu_l}(\delta T_B, \delta T_D)$, which should be given by the sum of the influence of drifts between evacuations, thus given by Eq. (51), and that during an evacuation, given by Eq. (34), can be written as

$$\bar{\delta}_{\Delta \nu_l}(\delta T_B, \delta T_D) = \left[ \frac{1}{3T_0} \frac{B_R(T_0)}{A_R^2} (n_i - 1)(T_0) + \frac{\varepsilon(T_0)}{T_0} - \alpha \right] \delta T_B$$
$$- \frac{\alpha}{n_i - 1} \delta T_D \quad . \quad (52)$$

The relative change in the frequency of the cavity mode addressed by the laser light in an empty cavity, i.e. $\bar{\delta}_{\nu_0}(\delta T_B, \delta T_D)$, is, as before, given by Eq. (S6) in the supplementary material.

As was done for the case with refractivity, when gas density is assessed by Eq. (47), using the assessed values of the shift of the beat frequency, $\Delta \nu_l(T_0, \delta T_B, \delta T_D)$, and the frequency of the cavity mode addressed, $\nu_0(T_0, \delta T_B, \delta T_D)$, according to the Eqs. (40) and (41), together with the characterized values of the $A_R(T_0)$, $\tilde{\chi}_{\Delta \nu}(T_0)$, and $\tilde{B}_{\Delta \nu}(T_0)$ - parameters, such a measurements will provide a value for the gas density, $(\rho_{n,i})_{T_0, \delta T_B, \delta T_D}$, that will differ from the actual one, $(\rho_{n,i})_{T_0}$, by an amount $\delta_{\rho_{n,i}}(\delta T_B, \delta T_D)$, and thereby give rise to a relative error, $\bar{\delta}_{\rho_{n,i}}(\delta T_B, \delta T_D)$, defined as $\delta_{\rho_{n,i}}(\delta T_B, \delta T_D)/\rho_{n,i}$, due to a change in the temperature during evacuation according to

$$(\rho_{n,i})_{T_0, \delta T_B, \delta T_D} = (\rho_{n,i})_{T_0} \left[ 1 + \bar{\delta}_{\rho_{n,i}}(\delta T_B, \delta T_D) \right]. \quad (53)$$

It can be shown that the relative error in the gas density, $\bar{\delta}_{\rho_{n,i}}(\delta T_B, \delta T_D)$, can be given by the same entities as the relative error in refractivity, $\bar{\delta}_{n_i-1}(\delta T_B, \delta T_D)$, viz. by Eq. (44), i.e. as the difference between the relative change in the shift of the beat frequency due to the changes in temperature, $\bar{\delta}_{\Delta \nu_l}(\delta T_B, \delta T_D)$, and the relative change in the frequency of the cavity mode addressed by the laser light in an empty cavity $\bar{\delta}_{\nu_0}(\delta T_B, \delta T_D)$. It is shown in the supplementary material, by Eq. (S15), that this give rise to a relative error of the assessment of the gas density, $\bar{\delta}_{\rho_{n,i}}(\delta T_B, \delta T_D)$, can be expressed as

$$\bar{\delta}_{\rho_{n,i}}(\delta T_B, \delta T_D) = \frac{\delta_{\rho_{n,i}}(\delta T_B, \delta T_D)}{\rho_{n,i}}$$
$$= \left[ \frac{1}{3T_0} \frac{B_R(T_0)}{A_R^2} (n_i - 1)(T_0) + \frac{\varepsilon(T_0)}{T_0} \right] \delta T_B - \frac{\alpha}{n_i - 1} \delta T_D. \quad (54)$$

Since $B_R(T_0)/A_R^2$ is around 0.032, this implies that the first term in the brackets takes, under STP conditions, a value of around $10^{-8}$. It was also concluded above that the contribution from the alteration of the instantaneous change in length of the cavity due to the presence of the gas in the cavity, given by $\varepsilon(T_0)/T_0$, will take values in the low $10^{-6}$ K$^{-1}$ to the low $10^{-5}$ K$^{-1}$ range (with the lower value for a well-constructed closed cavity and the higher for an open one). This implies that the contribution from the second virial coefficient, and thereby the temperature dependence of the refractivity of a given density, can be neglected with respect to the instantaneous change in length of the cavity. This implies that the temperature dependence of an assessment of density, on a relative scale, is the same as that of refractivity, viz. given by Eq. (46). Hence, it is significantly smaller for drifts in temperature between evacuations ($\delta T_B$) than during ($\delta T_D$); about a factor of three for the case with an open cavity made of ULE and about a factor of 300 for a closed cavity made of Zerodur.

## VI. MEANS TO CHARACTERIZE A FS-DFCB-OR SYSTEM

As was concluded above, the accuracy of the FS-DFCB-OR methodologies can be improved if the systems can be characterized with respect to some references or standards. Although presently there are no highly accurate refractivity or gas density standards available, it is of interest to scrutinize to which degree the accuracy of the FS-DFCB-OR methodologies could be improved if such would be available, or to which extent the methodologies can rely on internal standards. We will therefore, henceforth, distinguish between *external (or absolute) accuracy* (when the system is calibrated with respect to an external traceable reference or standard) and ***internal accuracy*** (when characterized with respect to an internal standard).

### A. Suitable means to characterize a FS-DFCB-OR system for assessment of refractivity

#### 1. Characterization by the use of a refractivity standard

If the system could be characterized by a gas whose refractivity is well-known under the conditions the system is to be used, the system would demonstrate a significantly better accuracy for assessment of refractivity than what was indicated in section III.B.2. above. As can be seen from Eq. (1), for the case with no relocking, it is possible to write the expression for the refractivity of the gas as

$$n_i - 1 = \Omega \frac{\Delta v_l}{v_0}\left[1 + \Omega \frac{\Delta v_l}{v_0} + \left(\frac{\Delta v_l}{v_0}\right)^2\right] \\ + (n_f - 1)\left[1 + 2\Omega \frac{\Delta v_l}{v_0}\right], \quad (55)$$

where a presumably negligible higher-order term has been neglected.

Since the estimates presented in Table 2 shows that $n_f - 1$ can be determined with such low uncertainty ($3 \times 10^{-12}$) that it only contributes to the relative uncertainty in the assessment of the refractivity (i.e. $\bar{\delta}_{n_i-1}$) with $1 \times 10^{-8}$, Eq. (55) shows that the system could be characterized for a given gas, by the use of a standard reference refractivity source, for a given set of conditions, down to uncertainty levels in this range, by assessing solely one parameter, *viz.* the $\Omega$ entity. This implies that, in this case, it is sufficient to characterize the system at a single refractivity. This is advantageous since, if standard reference refractivity sources with good accuracy could be produced, they do not need to be available at a variety of refractivities; it is sufficient if it could be produced for a single set of conditions. This thus indicates that, if a standard reference refractivity source would be available at least at one refractivity, it should be possible to assess the $\Omega$ entity, and thereby the entire FS-DFCB-OR system for assessment of refractivity, with that single refractivity characterization, with a relative uncertainty that is equal to that of the reference source, down to a level given by that of the residual refractivity.

However, if standard reference refractivity sources could be produced at a variety of refractivities, it is possible (and even advisory) to characterize the instrument at several refractivities. Such a multi-refractivity characterization could improve on the characterization of the system, not only for the cases when the system is affected by experimental uncertainties or errors but also for the cases when the uncertainly in the standard reference refractivity source is smaller (better) than that of the assessment of the refractivity, since then it possible to improve on also the uncertainty of the latter.

If a characterization of either of these two types could be performed it would thus be possible to obtain a system whose (relative) uncertainty is equal to that of the standard reference refractivity source.

However, since the parameters that make up $\Omega$, i.e. $\varepsilon$, $\varsigma$, and $\eta$, depend on one or several of a variety of physical entities,[27] if a system is to be used under conditions that differ from those at which it is characterized, a more elaborate characterization procedure needs to be used. However, this is a complex task that needs a detailed analysis. Although the temperature dependence of a typical FS-DFCB-OR instrumentation was scrutinized in some detail above, because of space constraints, this issue is relegated to a future publication.

#### 2. Requirements on standard reference refractivity sources for FS-DFCB-OR

In the lack of standard reference refractivity sources, one means to benefit from the extraordinary properties of the DF-(D)FPC-OR technique is to use a well characterized gas under well-specified conditions (in principle, for a given set of $\varepsilon$, $\varsigma$, and $\eta$) as a basis for a standard and perform assessments relative to this. In order to realize such a standard, at least a couple of prerequisites are needed:

*(i)* the gas used for the refractivity standard needs to have its refractivity determined or assessed (theoretically or experimentally) with high accuracy under at least one set of well characterized conditions; and

*(ii)* at the time of the system characterization, the gas must be presented under the same conditions at which its refractivity once was assessed.

If the latter cannot be obtained, e.g. if the gas only is accessible under dissimilar conditions (e.g. at a gas density that differs from that at which it was once characterized), then an alternative prerequisite needs to be fulfilled, *viz.*

*(iii)* the relation between refractivity and gas density must be known with sufficient accuracy.

The first prerequisite calls for a gas whose refractivity is well known. However, since there is a lack of refractivity standards, this is therefore not trivial to achieve. The second prerequisite requires means to present gas density with low uncertainty. Although one possible means to realize this is to measure the pressure of the gas, this requires not only a pressure measuring system with sufficient high accuracy but also good knowledge about the equation of state (i.e. how $p$ is related to $\rho$ for a given $T$) of the gas addressed. Although work to realize this is being pursued at various places, both these are difficult to realize with the highest accuracy. Therefore, also this is non-trivial to realize with high accuracy.

All this shows that the external accuracy of OR methodologies is presently limited by a lack of suitable refractivity standard. Work on how to circumvent this therefore takes place at several places.

Despite this, while awaiting the realization of suitable standard reference refractivity sources, so as to improve on the performance of OR techniques, various OR instrumentations in general, and DF- and FS-DFPC-OS systems in particular, can be developed for improved performance regarding other aspects.

#### 3. The use of an internal standard for assessment of refractivity by FS-DFCB-OR

One way to proceed is to perform assessments of refractivity that are relative to a well-defined experimental situation, i.e. to define an "OR internal refractivity standard" (OR-IRS), to which the refractivities of other gases are to be assessed and to which the system can be characterized. One possible means to realize this would be to address a given gas under well-defined conditions, predominantly at a given well-defined temperature [e.g. the melting point of Ga (29.7646 °C) is a defining thermometric fixed point of the International Temperature Scale of 1990 (ITS-90)[28, 29]. As long as measurements are performed under conditions that are similar to those of the characterization, it is possible to obtain refractivity assessments that have uncertainties relative to that of the OR-IRS that are below the external (or absolute) accuracy of the OR-IRS. The refractivity of this OR-IRS might be assessed on an absolute scale at some time in the future whereby also all up-to-then performed assessments will be assessed on an absolute scale. Although it is a challenge to develop such an OR-IRS, to benefit the most from OR techniques in general, and the DF-(D)FPC-OR technique and the FS-DFCB-OR methodology in particular, this is an important enterprise that should be pursued by the community in the closest future.

It needs to be emphasized though that for the case when a given system is run under conditions that are dissimilar from those at which it was characterized, additional uncertainties, originating from the uncertainty of the model used to relate the various entities to each other and uncertainties of the individual values of the various parameters when assessed in a characterization process, can also contribute to the final accuracy of the instrumentation. To reduce the risk for excessive uncertainties, detailed descriptions of the technique, incorporating all possible phenomena that can affect any such assessment, are therefore urgently needed. Such expressions, rigorously and meticulously derived, are therefore presented in our accompanying work.[14] Hence, these are of particular importance for FS-DFPC-OR instrumentations.

## B. Suitable means to characterize a FS-DFCB-OR system for assessment of gas density

### 1. Characterization by the use of a gas density standard

As for refractivity, if the system could be characterized by one type of gas whose density is well-know, the system would demonstrate a significantly better accuracy for assessment of gas density than when used in an uncharacterized mode of operation.

Although it is in general recommended that each physical entity that affects an assessment, e.g. the polarizability, $A_R$, the instantaneous deformation of the cavity, $\varepsilon$, and the gas and mirror dispersions, $\varsigma$ and $\eta$, should be assessed separately, it should also be noted that, as long as the system is used under such conditions that some of these vary simultaneously, some entities can alternatively be assessed as a group, as was the case for the $\varepsilon$, $\varsigma$, and $\eta$ entities that jointly make up the $\Omega(\varepsilon,\varsigma,\eta)$ entity in the expression for refractivity above [Eq. (55)].

For the case with gas density, it is possible to conclude that as long as the system is characterized in a way similar to as it later will be used, and for the cases when the residual gas density varies sufficiently little between the characterizations and the assessments so that the $\tilde{\chi}_{\Delta\nu}$ entity, which comprises $\Omega(\varepsilon,\varsigma,\eta)$ and a term proportional to $n_f - 1$, can be considered to be constant, it is possible to write the expression for the assessment of the density of a gas, for the case with no relocking, as a 3:rd order polynomial, i.e. as

$$\rho_{n,i} = \rho_{n,f} + C_1 \frac{\Delta\nu_l}{\nu_0}\left[1 + \tilde{B}_{\Delta\nu}\frac{\Delta\nu_l}{\nu_0} + \tilde{C}_{\Delta\nu}\left(\frac{\Delta\nu_l}{\nu_0}\right)^2\right], \quad (55)$$

where $C_1$ is a combined entity that is given by $(2/3A_R)\tilde{\chi}_{\Delta\nu}$.

This implies that a characterization of an FS-DFCB-OR instrumentation requires three coefficients to be determined, $C_1$, $\tilde{B}_{\Delta\nu}$, and $\tilde{C}_{\Delta\nu}$. This reduces significantly the complexity of the characterization. A characterization using at least three densities, but preferable more, would then provide all necessary information about the system so that it can be used for accurate assessments of gas densities under similar types of conditions as it was characterized (primarily the same temperature and wavelength but possibly dissimilar $\rho_{n,i}$).

### 2. Realization of a gas density standard

The most straightforward way to produce a standard reference gas density is to assess it from the pressure and temperature through an equation of state of the gas. To obtain the best accuracy, it is advisory to utilize a temperature stabilization that is based upon a stable temperature source, e.g., as was alluded to above, the melting point of Ga. Regarding assessment or control of the pressure, it is recommended to use of a high accuracy pressure gauge, e.g. a Gas Piston Gauge (GPG) such as RUSKA 2465A-754 (Fluke Calibration), since such can produce pressures with an accuracy of 10 ppm or 0.07 Pa, whichever is greater. Under the condition that the equation of state of the gas is sufficiently well-known (with an accuracy of at least $10^{-5}$), it would then be possible to assess all parameters in Eq. (55) with good accuracy, in particular the $C_1$ parameter with the same accuracy as that of the GPG, and the other two with sufficient accuracy. This would provide a reasonably adequate standard reference source of gas density with an accuracy of at least $10^{-5}$ over a large range of densities.

Alternatively, characterization could be performed by the use of a gas whose polarizability can be calculated with good accuracy. Of special importance in this case is He since its static polarizability has been calculated with an uncertainty of 0.1 ppm.[30, 31] Following the procedure presented by Stone et al.[6] it is possible, from this knowledge, to calculate the $A_R$, $\tilde{B}_{\Delta\nu}$, and $\tilde{C}_{\Delta\nu}$ coefficients with good accuracy; the former one with a relative uncertainty, $\delta A_R$, of $10^{-6}$ (and with the other two with "sufficiently good" uncertainties, $\delta_{\tilde{B}_{\Delta\nu}}$ and $\delta_{\tilde{C}_{\Delta\nu}}$, so their contributions to the total accuracy of the assessment can be neglected). By this, it is possible to determine the cavity deformation parameter, $\varepsilon$, with an accuracy that is limited by either $\delta A_R$ or the pressure standard used (whatever is largest). A drawback of this though is that the index of refraction of He is approximately one order of magnitude times lower than that for nitrogen, which can affect the possibility to perform the required assessments for low gas densities with the required accuracy.

### 3. The use of an internal standard for assessment of gas density by FS-DFCB-OR

For the case the equation of state of the gas would not be known with sufficient accuracy, it is possible to use the aforementioned system with liquid Ga as an internal standard to which gas densities can be assessed. Such should be made according to the prescription for the external gas density standard given above. It is even of interest to realize and define such a standard as an internal OR standard in such a way that it can be used by the entire community of users of DF-DFPC-OR under given conditions. As for refractivity, assessments of gas density should then be made with respect to these until sufficiently accurate gas density standards are developed.

## VII. SUMMARY AND CONCLUSIONS

Since the density of a gas in a finite volume is independent of temperature, quantification of amounts of gases can be done more accurately by assessment of the gas density than pressure. Moreover, refractivity is a function of density, which, to leading order, also is independent of temperature. Hence, since optical refractometry (OR) techniques measure refractivity, they have some distinct and indisputable advantages over pressure measuring techniques when it comes to quantification of amounts of gases. This implies that this type of technique has a great potential for assessment of gas density, and changes in such, under variety of conditions.

Unfortunately, however, the assessment of refractivity by OR is not always independent of temperature, mainly due to thermal drifts of the interferometer; in the case of a FP cavity, primarily by the material that constitutes the spacer between the mirrors. One means to minimize these types of effects is to use two FP cavities often referred to as a Dual Fabry-Perot Cavity (DFPC), here denoted DFPC-OR and to perform measurements under conditions that minimize the influence of such effects. Despite this, it has though been found that also when DFPC-OR is performed, the instrumentation is affected by drifts and relaxations in the cavity material.

To alleviate these types of limitations, and to contribute to the advancement of the FPC-OR technique, we present, in a series of papers, novel methodologies to assess both the gas density, and changes in such, that minimize the influence of the aforementioned limitations by measurements under drift free conditions.[13, 14]

In one of these,[13] we predict that DF-DFPC-OR can be realized by performing measurements on "short" time scales. By designing methodologies that are based on this principle, which can be made by rapidly switching gas volumes, referred to as *fast switching* FPC-OR and denoted FS-FPC-OR (or FS-DFPC-OR), it is prophesized that OR can be utilized for high precision and accurate assessments of gas density under drift-free cavity conditions, i.e. virtually unaffected by some of the limitations imposed by the cavity spacer material, and hence closer to its full power. An important advantage of this is that such methodologies could then be constructed around systems not exposed to elaborate baking procedures.

In addition, to benefit the most from this, we present, in Ref. [14], derivations of proper relations between the refractivity as well as the density of the gas in a cavity and the measured change in frequency of laser light that is locked to a mode of a cavity as the cavity is evacuated. Since this is done by properly acknowledging the influence of various higher order and non-linear contributions to these relations, including the effect of the deformation of the cavity due to the presence of the gas and the dispersion of the mirrors., these expressions are assumed to be the basis for highly accurate assessments of refractivity and gas density by FPC-OR techniques.

Based on these two works we have in this work critically assessed the expected precision and accuracy of FS-DFPC-OR. The suggested methodologies have been scrutinized with respect to a number of properties; first regarding their ability to assess refractivity, and then gas density. For each of these, first the precision and then the accuracy were estimated. After this, the temperature dependence of the methodologies was assessed.

It was concluded that the FS-DFCB-OR methodology can provide a precision that is independent of the gas density, for the pertinent condition a precision in refractivity of $5 \times 10^{-13}$ and in gas density of $9 \times 10^{-8}$ mol/m$^3$. This implies that it has its highest resolution for the highest densities; the relative precisions in both refractivity and gas density under STP conditions are $2 \times 10^{-9}$, which is several orders of magnitude better than the relative accuracy. This preponderance to precision originates from the fact that the technique converts a change in refractivity (and density) to frequency, which is an entity that can be assessed with exceptionally high precision. This opens up for the use of the technique under conditions when a high resolution is required, e.g. for assessment of differences or changes of refractivity and gas density under a variety of conditions, in particular those that are not temperature stabilized. As is further alluded to in one of our accompanying paper,[13] since assessments of small changes in gas density, e.g. those from leaks, relies more on a high precision than a high accuracy, this shows that the technique is particularly suitable for assessments of changes in gas density.

One reason for the significantly poorer accuracy than precision is that there are currently no accurate sources of calibration, neither for refractivity, nor for gas density. It was discussed though that, under some general conditions, the system can be fully characterized for assessment of both refractivity and gas density if solely a single standard reference refractivity or gas density source would be available. It was also concluded, as an alternative, that the system could be characterized for its ability to assess refractivity and gas density by the use of a suitable internal standard. All this opens up for highly accurate use the methodology in the future.

It was also shown that the FS-DFCB-OR technique has an extremely small temperature dependence; in fact, several orders of magnitude smaller than that of pressure measuring devices; the influence of temperature drifts during an evacuation is mainly given by the thermal expansion of the cavity spacer, while that between evacuations is given by the (temperature dependence of the) instantaneous length deformation of the cavity. It was demonstrated that the methodologies makes the system significantly less sensitive to temperature changes between evacuations than during. Since an evacuation can be made swiftly, i.e. in the orders of seconds, for which the temperature drift can be assumed to be small, the methodology has an exceptionally small temperature dependence. It was moreover concluded that a drift *between* evacuations can be seen as a *reduction of the accuracy* of the technique while a drift *during* evacuations can be seen as a *reduction of the precision* or the resolution of the technique.

By the analysis presented in work, together with our two accompanying works, Ref [14], in which we provide the basis of the DF-DFPC-OR technique and the principles of the FS-DFCB-OR methodologies by deriving explicit expressions for how the refractivity and gas density can be assessed by assessments of changes in beat frequency between two laser fields, and Ref. [13], in which a possible realization of the FS-DFCB-OR methodology is described, we claim that we have laid the foundation for a methodology for FPC-OR in general, and DFPC-OR in particular, that are based on rapid changes of gas densities in the measurement cavity, termed fast switching DFCB-OR (FS-DFCB-OR), that can be used for precise and accurate assessment of gas density, with a precision that is significantly better than the accuracy, under a variety of conditions, including the important non-temperature stabilized ones. Focusing on the applications when the extraordinary precision and low temperature dependence come to their right, and by defining and using an internal standard, possibly referred to as an OR internal refractivity standard" (OR-IRS), the present lack of absolute refractivity standards should not restrict the further development of these types of instrumentation. It is hoped that this work can serve as a basis for future constructions of instrumentations that can benefit the most from the extraordinary power of DFPC-OR for refractivity and gas density assessments.

## SUPPLEMENTARY MATERIAL

See the supplementary material for information about the influence of a temperature drift on the assessment of refractivity by FS-DFCB-OR (part 1); the influence of temperature drift on the frequency of the cavity mode addressed by the laser light in an empty cavity (part 2); and the influence of a temperature drift on the assessment of gas density by FS-DFCB-OR (part 3).

## Acknowledgment

This research was supported by the EMPIR initiative, which is co-founded by the European Union's Horizon 2020 research and innovation programme and the EMPIR Participating States; the Swedish Research Council (VR), project numbers 621-2011-4216 and 621-2015-04374; Umeå University program VFS (Verifiering för samverkan), 2016:01; and Vinnova Metrology Programme, project number 2015-0647 and 2014-06095.

[15] It is plausible to presume that, for a series of consecutive evacuations, for the case with no relocking, it is possible to keep the laser locked to the same mode all the time, while, for the case with relocking, to repeatedly relock the laser to the same mode. This implies that $\Delta q$ and $q_0$ can be assumed to be fixed entities also for the case with multiple gas evacuation measurements, whereby the $\sigma_{\Delta q}$ and $\sigma_{q0}$ entities can be neglected also when the MGE precision is estimated.

[16] In the case when $A_R$ is given in units of m$^3$, the molar gas constant, $R$, in Eq. (10) should be exchanged to the Boltzmann constant, $k$.

[17] The reason for this is that the pressure gauge does not have to have any response at $p_i$; it can be selected for having the best performance for the residual pressure in the cell. Hence, by choosing a pressure gauge whose maximum pressure is slightly above $p_f$, the relative fluctuations of the residual refractivity, given by Eq. (11), can therefore in principle be several orders of magnitude below the resolution of the pressure gauge.

[20] Note here that although the methodology presented is valid for any wavelength of the light, the estimates presented here are based upon the same wavelength that was used in the experimental work described in our accompanying paper, i.e. at 1.5 μm.

[23] As is discussed in our accompanying paper,[14] since the cavity spacer is made of a low thermal expansion material, α is in the order of 10$^{-8}$ or 10$^{-7}$ K$^{-1}$,[3] where the lower value is valid for the best types of ultra-low expansion (ULE) glass. This implies that, under STP conditions, for which $n$-1 is around $3 \times 10^{-4}$, the $\alpha/(n-1)$ term takes typically values in the order of $3 \times 10^{-5}$ to $3 \times 10^{-4}$ K$^{-1}$, where the lower value is valid for the best types of ULE glass. It was also argued in the same reference that $\varepsilon$ can have a value in the low 10$^{-4}$ range for a well-constructed closed cavity (although it typically takes a value in the low 10$^{-3}$ range for an open cavity). This implies that, under STP conditions, the $\varepsilon(T_0)/T_0$ term will take values in the low 10$^{-6}$ K$^{-1}$ to the low 10$^{-5}$ K$^{-1}$ range, which thus are 3 to 300 times smaller than those of $\alpha/(n-1)$.

[24] This expression can alternatively be obtained by differentiating Eq. (37).

[25] Since the last term in the second set of parentheses in Eq. (45) originates from the fact that $v_0(T_0+\delta T)$ differs from $v_0(T_0)$, this indicates that the refractivity is not in practice influenced by whether we monitor $v_0(T_0)$ or $v_0(T_0+\delta T)$.

[26] It suffices to evacuate the cavity down to 1 mTorr to produce a final density, $\rho_{n,f}$, that is in the order of 10$^{-6}$ of that of $\rho_{n,i}$ at STP conditions. A drift in temperature of 1 mK implies that this residual density term contributes to the temperature dependence of $\rho_{n,i}$ on the order of $3 \times 10^{-12}$, which is significantly smaller than the contributions from other effects.

[27] Primarily the frequency or wavelength of the light ($v_0$ or $\lambda$), the length of the cavity, $L_0$, some specific properties of the cavity material (e.g. the Young's modulus, $E$, and the geometrical design of the cavity, $\beta$), some properties of the mirror material (e.g. the Poisson's ratio, $v(bar)$, the shear modulus, $G$, and the thickness of the mirror substrate, $h$), the dispersion of the mirrors and the gas, $GDD(v_c)$ and η, respectively, the molar refractivity of the gas, $A_R$, and the temperature, $T$, all properly defined in the supplementary material of Ref. 14.

# Supplementary material to "Fast switching dual Fabry-Perot-cavity-based optical refractometry for assessment of gas refractivity and density – estimates of its precision, accuracy, and temperature dependence"

Martin Zelan,[1,a)] Isak Silander,[2] Thomas Hausmaninger,[2] and Ove Axner[2,a)]

[1] Measurement Science and Technology, RISE Research Institutes of Sweden, SE-501 15 Borås, Sweden
[2] Department of Physics, Umeå University, SE-901 87 Umeå, Sweden

**Part 1 – Influence of a temperature drift on the assessment of refractivity by FS-DFCB-OR**

It was shown by Eq. (44) that for the case when the temperature drifts both between and during evacuations, i.e. from $T_0$ to $T_0 + \delta T_B + \delta T_D$, the relative error in the assessment of the refractivity, $\bar{\delta}_{n_i-1}(\delta T_B, \delta T_D)$, is given by $\bar{\delta}_{\Delta \nu_l}(\delta T_B, \delta T_D) - \bar{\delta}_{\nu_0}(\delta T_B, \delta T_D)$. In order to asses these entities, let us assess separately the change in $\Delta \nu_l$ when the temperature drifts between evacuations of the cavity an amount $\delta T_B$, from $T_0$ (at the time of the characterization) to $T_0 + \delta T_B$ (at the time of the measurement), denoted, $\delta_{\Delta \nu_l}(\delta T_B)$, and when the temperature drifts during the evacuation, from a temperature of $T$ at the time the measurement starts, to a temperature of $T + \delta T_D$ at the end of the evacuation, denoted $\delta_{\Delta \nu_l}(\delta T_D)$. Let us thereafter estimate how a combined drift of temperature, from $T_0$ to $T_0 + \delta T_B + \delta T_D$, affects the relative change in the frequency of the cavity mode addressed by the laser light in an empty cavity, $\bar{\delta}_{\nu_0}(\delta T_B, \delta T_D)$.

*Part 1.1. Influence of a temperature drift between evacuations of the cavity for a temperature independent refractivity*

Following the derivation of the expression for the shift in the cavity mode frequency that takes place during an evacuation, which, in the absence of relocking, is equal to the shift in the laser beat frequency, $\Delta \nu_l$, (given in the supplementary material our accompanying work [1]), it is possible to express the change in $\Delta \nu_l$ due to a temperature drift between evacuations of the cavity an amount $\delta T_B$ [under the condition that temperature has changed from $T_0$ (at the time of the characterization) to $T_0 + \delta T_B$ (at the time of the measurement)], denoted $\delta_{\Delta \nu_l}(\delta T_B)$, from a fixed gas refractivity [for which it can be assumed that $(n)_{T_0} = (n)_{T_0 + \delta T_B}$], as

$$\delta_{\Delta \nu_l}(\delta T_B) = \left\{ [\nu_q(n_f)] - [\nu_q(n_i)] \right\}_{T_0 + \delta T_B}$$
$$- \left\{ [\nu_q(n_f)] - [\nu_q(n_i)] \right\}_{T_0}$$
$$= \left( \frac{qc}{2n_f L} \right)_{T_0 + \delta T_B} - \left( \frac{qc}{2n_i L} \right)_{T_0 + \delta T_B} - \left( \frac{qc}{2n_f L} \right)_{T_0} + \left( \frac{qc}{2n_i L} \right)_{T_0}$$
$$= \frac{qc}{2n_f(T_0)L_0(T_0)} \left\{ \frac{1}{1 + \alpha \delta T_B + \delta[\Delta \bar{L}_E(n_f), \delta T_B]} - 1 \right\}$$
$$- \frac{qc}{2n_i(T_0)L_0(T_0)} \left\{ \frac{1}{1 + \alpha \delta T_B + \delta[\Delta \bar{L}_E(n_i), \delta T_B]} - 1 \right\}$$
$$\approx \frac{\nu_0(T_0)}{n_i(T_0)} \left\{ \alpha \delta T_B + \delta[\Delta \bar{L}_E(n_i), \delta T_B] \right\}$$
$$- \frac{\nu_0(T_0)}{n_f(T_0)} \left\{ \alpha \delta T_B + \delta[\Delta \bar{L}_E(n_f), \delta T_B] \right\}$$

$$= \nu_0(T_0) \left[ \left( \frac{1}{n_i(T_0)} - \frac{1}{n_f(T_0)} \right) \alpha \delta T_B \right.$$
$$\left. + \left( \frac{n_i(T_0) - 1}{n_i(T_0)} - \frac{n_f(T_0) - 1}{n_f(T_0)} \right) \delta \varepsilon(\delta T_B) \right] \quad \text{(S1)}$$
$$= \nu_0(T_0) \frac{n_i(T_0) - n_f(T_0)}{n_f(T_0)n_i(T_0)} \left[ \frac{\varepsilon(T_0)}{T_0} - \alpha \right] \delta T_B,$$

where $L_0(T_0)$ is the length of the empty cavity at the temperature $T_0$, $\alpha$ is the thermal expansion coefficient of the spacer material, $\delta[\Delta \bar{L}_E(n), \delta T_B]$ is the change in $\Delta \bar{L}_E(n)$ due to the change of the temperature, where $\Delta \bar{L}_E(n)$ the instantaneous change in length of the cavity to due to the presence of the residual gas for a gas refractivity of $n-1$ {which is given by Eq. (S22) in our accompanying paper [1]}, $\nu_0(T_0)$ is the frequency of the cavity mode addressed in an empty cavity at a temperature of $T_0$, given by $qc/[2L_0(T_0)]$, and $\delta \varepsilon(\delta T_B)$ represents the drift in the $\varepsilon$-parameter due to the drift in temperature. We have also, in the last step, made use of the fact that $\varepsilon$ is proportional to temperature (which originates from its dependence on pressure), whereby we have written $\delta \varepsilon(\delta T_B)$ as $\varepsilon(T_0) \cdot (\delta T_B / T_0)$.

Since $\Delta \nu_l$ is given by $(n_i - n_f)\nu_0(T_0)$ to first order, this implies that the relative change in the shift of the laser beat frequency, denoted $\bar{\delta}_{\Delta \nu_l}(\delta T_B)$ and defined as $\delta_{\Delta \nu_l}(\delta T_B) / \Delta \nu_l$, can be expressed as

$$\bar{\delta}_{\Delta \nu_l}(\delta T_B) = \frac{\delta_{\Delta \nu_l}(\delta T_B)}{\Delta \nu_l} = \left[ \frac{\varepsilon(T_0)}{T_0} - \alpha \right] \delta T_B, \quad \text{(S2)}$$

where we have assumed that $n_i n_f$ can be approximated by unity.

*Part 1.2. Influence of a temperature drift during evacuations of the cavity for a temperature independent refractivity*

For the case when the temperature drifts during the evacuation, from a temperature of $T$ at the time the measurement starts, to a temperature of $T + \delta T_D$ at the end of the evacuation, the change of the shift in the laser beat frequency (thus due to a change in temperature of $\delta T_D$), denoted, $\delta_{\Delta \nu_l}(\delta T_D)$, can be expressed as

$$\delta_{\Delta \nu_l}(\delta T_D) = \left\{ [\nu_q(n_f)]_{T+\delta T_D} - [\nu_q(n_i)]_T \right\}$$
$$- \left\{ [\nu_q(n_f)]_T - [\nu_q(n_i)]_T \right\}$$
$$= [\nu_q(n_f)]_{T+\delta T_D} - [\nu_q(n_f)]_T$$
$$= \left( \frac{qc}{2n_f L} \right)_{T+\delta T_D} - \left( \frac{qc}{2n_f L} \right)_T$$

$$= \frac{qc}{2n_f(T)L_0(T)\left[1+\Delta\bar{L}_E(n_f)\right]}$$
$$\times\left\{\frac{1}{1+\alpha\delta T_D+\delta[\Delta\bar{L}_E(n_f),\delta T_D]}-1\right\}$$
$$\approx -\frac{\nu_0(T)}{n_f(T)}\left[1-\Delta\bar{L}_E(n_f)\right]\left\{\alpha\delta T_D+\delta[\Delta\bar{L}_E(n_f),\delta T_D]\right\} \quad (S3)$$
$$= -\nu_0(T)\left[1-(n_f-1)\varepsilon(T)\right]\left[\alpha\delta T_D+(n_f-1)\delta\varepsilon(\delta T_D)\right]$$
$$= -\nu_0(T)\left[\alpha+(n_f-1)\frac{\varepsilon(T)}{T}\right]\delta T_D$$
$$\approx -\nu_0(T)\alpha\delta T_D,$$

where we in the last step have utilized the fact that the entire $\varepsilon$-terms is small, primarily due to the smallness of both $\varepsilon$ and $n_f-1$.

Since the shift in the laser frequency, $\Delta\nu_l$, can be expressed as $(n_i-n_f)\nu_0(T)$, this implies that the relative change in the shift in the laser frequency, denoted $\bar{\delta}_{\Delta\nu_l}(\delta T_D)$ and defined as $\delta_{\Delta\nu_l}(\delta T_D)/\Delta\nu_l$, can be expressed alternatively as

$$\bar{\delta}_{\Delta\nu_l}(\delta T_D) = \frac{\delta_{\Delta\nu_l}(\delta T_D)}{\Delta\nu_l} = -\frac{\alpha}{n_i-n_f}\delta T_D = -\frac{2\alpha}{3A_R\rho_{n,i}}\delta T_D, \quad (S4)$$

where we in the last step have assumed that $n_i-n_f$ can be written as $3A_R(\rho_{n,i}-\rho_{n,f})/2$ and that $\rho_{n,i}\gg\rho_{n,f}$.

**Part 2 – Influence of temperature drift on the frequency of the cavity mode addressed by the laser light in an empty cavity**

The relative change in the frequency of the cavity mode addressed by the laser light in an empty cavity due to a drift of temperature from $T_0$ to $T_0+\delta T_B+\delta T_D$, $\bar{\delta}_{\nu_0}(\delta T_B,\delta T_D)$, can be assessed by first considering that the relative length of the cavity increases with temperature as $1+\alpha(\delta T_B+\delta T_D)+\delta[\Delta\bar{L}_E(n_i),\delta T_B] - \delta[\Delta\bar{L}_E(n_f),\delta T_B] + \delta[\Delta\bar{L}_E(n_f),\delta T_D]$, where the various $\delta[\Delta\bar{L}_E(n),\delta T]$ entities represent the changes in $\Delta\bar{L}_E(n)$ due to the change of the temperature $\delta T$. This implies that $\nu_0(T_0,\delta T_B,\delta T_D)$ can be written as

$$\nu_0(T_0,\delta T_B,\delta T_D)$$
$$= \nu_0(T_0)\frac{1}{1+\alpha(\delta T_B+\delta T_D)+\delta[\Delta\bar{L}_E(n_i),\delta T_B]}$$
$$\approx \nu_0(T_0)\left[1-\alpha(\delta T_D+\delta T_B)-(n_i-1)\delta\varepsilon(\delta T_B)\right] \quad (S5)$$
$$\approx \nu_0(T_0)\left[1-\alpha\delta T_D-\left[\alpha+(n_i-1)\frac{\varepsilon(T)}{T}\right]\delta T_B\right]$$

where we have used the fact that both $\delta[\Delta\bar{L}_E(n_f),\delta T_B]$ and $\delta[\Delta\bar{L}_E(n_f),\delta T_D]$ are much smaller than the other terms, whereby they have been neglected, and in the second step made use of the smallness of all entities and series expanded the expression. We will henceforth denote the change in the frequency of the cavity mode addressed by the laser light in an empty cavity by $\delta_{\nu_0}(\delta T_B,\delta T_D)$, defined by $\nu_0(T_0,\delta T_B,\delta T_D)-\nu_0(T_0)$, and its relative counterpart, denoted $\bar{\delta}_{\nu_0}(\delta T_B,\delta T_D)$ and defined as $\delta_{\nu_0}(\delta T_B,\delta T_D)/\nu_0(T_0)$. The expression above shows that

$$\bar{\delta}_{\nu_0}(\delta T_B,\delta T_D) = \frac{\delta_{\nu_0}(\delta T_B,\delta T_D)}{\nu_0(T_0)}$$
$$= -\left[\alpha+(n_i-1)\frac{\varepsilon(T_0)}{T_0}\right]\delta T_B-\alpha\delta T_D \quad (S6)$$
$$= -\alpha(\delta T_D+\delta T_B)-(n_i-1)\frac{\varepsilon(T_0)}{T_0}\delta T_B$$
$$\approx -\alpha(\delta T_B+\delta T_D)$$

where we in the last step have made use of the fact that $(n_i-1)[\varepsilon(T_0)/T_0]$, is typically in the $3\times10^{-10}$ to $3\times10^{-9}$ K$^{-1}$ range, while $\alpha$ is in the $10^{-8}$ to $10^{-7}$ K$^{-1}$ range, and thereby neglected the influence of the instantaneous change in length of the cavity to due to the presence of the gas in the cavity with respect to the change in length of the cavity due to thermal expansion of the spacer material.

**Part 3 – Influence of a temperature drift on the assessment of gas density by FS-DFCB-OR**

*Part 3.1. Influence of a temperature drift between evacuations of the cavity on the refractivity for a given, temperature independent gas density*

When the influence of temperature drifts on the assessment of gas density is to be assessed, it is most convenient to scrutinize the changes in the shift of the beat frequency of the light and the assessed gas density when a fixed gas density is considered. In this case, however, we have to be taken into account the fact that if the temperature changes between evacuations of the cavity, a constant gas density will give rise to dissimilar refractivities. This can be inferred from Eq. (S48) in our accompanying paper,[1] which shows that the refractivity of a gas with a given density depends on the temperature through the higher order virial coefficients. It was stated there that the refractivity can be written as

$$(n-1)(T) = A_m\rho_n+B_m(T)\rho_n^2+... \quad (S7)$$

where $B_m(T)$ is given by $3A_R^2/8+3B_R(T)/2$ where, in turn, $A_R$ and $B_R(T)$ are the first two virial coefficients in the extended Lorentz-Lorenz equation [Eq. (15) in the same reference]. Since $B_R(T)$ depends on temperature, this implies that we can no longer assume that refraction of the gas is independent of the temperature, i.e. that $(n)_{T_0}=(n)_{T_0+\delta T_B}$ and we need to consider the temperature dependence of the refractivity of a gas held at a fixed density.

When this is done, it is suitable to express the refractivity at a temperature of $T_0+\delta T_B$ in terms of that at the temperature of $T_0$ and the change in refractivity due to the change in temperature $\delta T_B$ as

$$(n-1)(T) = (n-1)(T_0)+\frac{\partial(n-1)(T)}{\partial T}\delta T_B$$
$$= (n-1)(T_0)+\frac{\partial B_m(T)}{\partial T}\rho_n^2\delta T_B$$
$$= (n-1)(T_0)\left\{1+\frac{1}{(n-1)(T_0)}\frac{\partial B_m(T)}{\partial T}\frac{2^2}{3^2 A_R^2}\left[(n-1)(T_0)\right]^2\delta T_B\right\} \quad (S8)$$
$$= (n-1)(T_0)\left\{1+\left[\frac{T_0}{B_R(T_0)}\frac{\partial B_R(T)}{\partial T}\right]\frac{2}{3T_0}\frac{B_R(T)}{A_R^2}\left[(n-1)(T_0)\right]\delta T_B\right\}$$

where we have used Eq. (S7) together with the definition of $B_m(T)$ and assumed that the $B_R(T)$ virial coefficient is the only temperature dependent entity. [2]

If we assume that $B_R(T)$ has a linear dependence on molecular velocity, which is reasonable,[3,4] and thereby a square root dependence on $T$, we get $[T_0/B_R(T_0)]\partial B_R(T)/\partial T=1/2$. This implies that

$$(n-1)(T) = (n-1)(T_0)\left\{1+\frac{1}{3T_0}\frac{B_R(T_0)}{A_R^2}\left[(n-1)(T_0)\right]\delta T_B\right\}$$
$$= (n-1)(T_0)\cdot\left\{1+\bar{\delta}[(n-1)(T_0),\delta T_B]\right\}, \quad (S9)$$

which shows that the relative change in the refractivity due to the change of the temperature, $\bar{\delta}[(n-1)(T_0),\delta T_B]$, is given by

$$\bar{\delta}[(n-1)(T_0), \delta T_B] = \frac{1}{3T_0} \frac{B_R(T_0)}{A_R^2} [(n-1)(T_0)] \delta T_B \quad \text{(S10)}$$

Since $B_R(T)/A_R^2$ is around 0.032, this implies that, under STP conditions,

$$\bar{\delta}[(n-1)(T_0), \delta T_B] = 1 \times 10^{-8} \cdot \delta T_B. \quad \text{(S11)}$$

*Part 3.2. Influence of a temperature drift between evacuations of the cavity on the index of refraction for a given, temperature independent gas density*

Similar to what was done for the refractivity, we can assess the influence of a temperature shift on the index of refraction. Following Eq. (S8), and making use of the fact that $[T_0/B_R(T_0)]\partial B_R(T)/\partial T = 1/2$, we can write

$$\begin{aligned}
n(T) &= n(T_0) + \frac{\partial n(T)}{\partial T} \delta T_B \\
&= n(T_0) + \frac{\partial B_m(T)}{\partial T} \rho_n^2 \delta T_B \\
&= n(T_0) \left\{ 1 + \frac{1}{3T_0} \frac{B_R(T)}{A_R^2} \frac{[(n-1)(T_0)]^2}{n(T_0)} \delta T_B \right\} \\
&= n(T_0) \left\{ 1 + \bar{\delta}[n(T_0), \delta T_B] \right\}
\end{aligned} \quad \text{(S12)}$$

*Part 3.3. Influence of a temperature drift between evacuations of the cavity on the shift of the beat frequency for a given, temperature independent gas density*

For the case when the temperature drifts between evacuations of the cavity, the change in the shift in the beat frequency due to a change in temperature of $\delta T_B$, denoted, $\delta_{\Delta v_l}(\delta T_B)$, can thereby be expressed as

$$\begin{aligned}
&\delta_{\Delta v_l}(\delta T_B) \\
&= \left\{ [v_q(n_f)] - [v_q(n_i)] \right\}_{T_0 + \delta T_B} \\
&\quad - \left\{ [v_q(n_f)] - [v_q(n_i)] \right\}_{T_0} \\
&= \left( \frac{qc}{2n_f L} \right)_{T_0 + \delta T_B} - \left( \frac{qc}{2n_i L} \right)_{T_0 + \delta T_B} - \left( \frac{qc}{2n_f L} \right)_{T_0} + \left( \frac{qc}{2n_i L} \right)_{T_0} \\
&= \frac{qc}{2n_f(T_0)L_0(T_0)} \\
&\quad \times \frac{1}{1+\bar{\delta}[n_f(T_0),\delta T_B]} \left\{ \frac{1}{1+\alpha\delta T_B + \delta[\Delta\bar{L}_E(n_f),\delta T_B]} - 1 \right\} \\
&\quad - \frac{qc}{2n_i(T_0)L_0(T_0)} \\
&\quad \times \frac{1}{1+\bar{\delta}[n_i(T_0),\delta T_B]} \left\{ \frac{1}{1+\alpha\delta T_B + \delta[\Delta\bar{L}_E(n_i),\delta T_B]} - 1 \right\} \\
&\approx \frac{v_0(T_0)}{n_i(T_0)} \left\{ \bar{\delta}[n_i(T_0),\delta T_B] + \alpha\delta T_B + \delta[\Delta\bar{L}_E(n_i),\delta T_B] \right\} \\
&\quad - \frac{v_0(T_0)}{n_f(T_0)} \left\{ \bar{\delta}[n_f(T_0),\delta T_B] + \alpha\delta T_B + \delta[\Delta\bar{L}_E(n_f),\delta T_B] \right\}
\end{aligned}$$

$$\begin{aligned}
&= v_0(T_0) \left\{ \frac{\bar{\delta}[n_i(T_0),\delta T_B]}{n_i(T_0)} - \frac{\bar{\delta}[n_f(T_0),\delta T_B]}{n_f(T_0)} \right. \\
&\quad + \left( \frac{1}{n_i(T_0)} - \frac{1}{n_f(T_0)} \right) \alpha \delta T_B \\
&\quad + \left. \left( \frac{n_i(T_0)-1}{n_i(T_0)} - \frac{n_f(T_0)-1}{n_f(T_0)} \right) \delta \varepsilon \right\} \\
&= v_0(T_0) \frac{n_i(T_0) - n_f(T_0)}{n_f(T_0)n_i(T_0)} \quad \text{(S13)} \\
&\quad \times \left\{ \frac{1}{3T_0} \frac{B_R(T_0)}{A_R^2} [(n_i-1)(T_0) + (n_f-1)(T_0)] + \frac{\varepsilon(T_0)}{T_0} - \alpha \right\} \delta T_B \\
&= v_0(T_0) [n_i(T_0) - n_f(T_0)] \\
&\quad \times \left\{ \frac{1}{3T_0} \frac{B_R(T_0)}{A_R^2} [(n_i-1)(T_0)] + \frac{\varepsilon(T_0)}{T_0} - \alpha \right\} \delta T_B,
\end{aligned}$$

where we in the last step have used the fact that $(n_f - 1)(T_0) \ll (n_i - 1)(T_0)$ and that $[n_f(T_0)n_i(T_0)]^{-1} \approx 1$.

This implies that the relative change in the shift in the beat frequency, denoted $\bar{\delta}_{\Delta v_l}(\delta T_B)$ and defined as $\delta_{\Delta v_l}(\delta T_B)/\Delta v_l$, can be expressed as

$$\begin{aligned}
\bar{\delta}_{\Delta v_l}(\delta T_B) &= \frac{\delta_{\Delta v_l}(\delta T_B)}{\Delta v_l} \\
&= \left\{ \frac{1}{3T_0} \frac{B_R(T_0)}{A_R^2} [(n_i-1)(T_0)] + \frac{\varepsilon(T_0)}{T_0} - \alpha \right\} \delta T_B
\end{aligned} \quad \text{(S14)}$$

*Part 3.4. Influence of a temperature drift between evacuations of the cavity on the assessment of gas density for a for a given, temperature independent gas density*

It can be shown that the relative error in the gas density, $\bar{\delta}_{\rho_{n,i}}(\delta T_B, \delta T_D)$, can be given by the same entities as the relative error in refractivity, $\bar{\delta}_{n_i-1}(\delta T_B, \delta T_D)$, viz. by Eq. (44), i.e. as the difference between the relative change in the shift of the beat frequency due to the changes in temperature, $\bar{\delta}_{\Delta v_l}(\delta T_B, \delta T_D)$, and the relative change in the frequency of the cavity mode addressed by the laser light in an empty cavity $\bar{\delta}_{v_0}(\delta T_B, \delta T_D)$. This implies that the relative error of the assessment of the gas density, $\bar{\delta}_{\rho_{n,i}}(\delta T_B, \delta T_D)$, can be expressed as

$$\begin{aligned}
\bar{\delta}_{\rho_{n,i}}(\delta T_B, \delta T_D) &= \frac{\delta_{\rho_{n,i}}(\delta T_B, \delta T_D)}{\rho_{n,i}} \\
&= \bar{\delta}_{\Delta v_l}(\delta T_B, \delta T_D) - \bar{\delta}_{v_0}(\delta T_B, \delta T_D) \\
&= \left\{ \frac{1}{3T_0} \frac{B_R(T_0)}{A_R^2} [(n_i-1)(T_0)] + \frac{\varepsilon(T_0)}{T_0} - \alpha \right\} \delta T_B \\
&\quad - \frac{\alpha}{n_i-1} \delta T_D + \alpha(\delta T_D + \delta T_B) \\
&= \left\{ \frac{1}{3T_0} \frac{B_R(T_0)}{A_R^2} [(n_i-1)(T_0)] + \frac{\varepsilon(T_0)}{T_0} \right\} \delta T_B \\
&\quad - \left[ \frac{\alpha}{n_i-1} - \alpha \right] \delta T_D \\
&\approx \left\{ \frac{1}{3T_0} \frac{B_R(T_0)}{A_R^2} [(n_i-1)(T_0)] + \frac{\varepsilon(T_0)}{T_0} \right\} \delta T_B \\
&\quad - \frac{\alpha}{n_i-1} \delta T_D
\end{aligned}$$

$$= \left\{ \frac{1}{3T_0} \frac{B_R(T_0)}{A_R^2} [(n_i - 1)(T_0)] + \frac{\varepsilon(T_0)}{T_0} \right\} \delta T_B$$
$$- \frac{2\alpha}{3 A_R \rho_{n,i}} \delta T_D \quad \text{(S15)}$$

where we, in the last steps, have made use of the fact that $(n_i - 1)^{-1} \gg 1$ and that $n_i - 1$ can be written as $3 A_R \rho_{n,i} / 2$.